\documentclass[tightenlines,aps,prb,amssymb,amsfonts,eqsecnum,showpacs,floatfix,twocolumn]{revtex4}

\usepackage{bm}
\usepackage{subfigure}
\usepackage{graphicx}

\newcommand{\sss}[1]{\bm{\left\{}\rule[-.751ex]{0ex}{0ex} #1\bm{\right\}}}
 \newcommand{\ssss}[1]{\bm{\left[} #1\bm{\right]}\hspace*{-.51ex}}
 \newcommand{\ssssd}[1]{\bm{\left[} #1\bm{\right]}^{\dagger}\hspace*{-.51ex}}

\begin{document}

\title{Quantum measurement in the charge representation}
\author{J. Rammer}
\affiliation{Department of Physics, Ume{\aa}  University, SE-901 87 Ume\aa}
\author{A. L.  Shelankov}
\altaffiliation[Also at ]{A. F. Ioffe Physico-Technical Institute, 19021 St. Petersburg, Russia.}
\affiliation{Department of Physics, Ume{\aa}  University, SE-901 87 Ume\aa}
\author{J. Wabnig}
\affiliation{Department of Physics, Ume{\aa}  University, SE-901 87 Ume\aa}
\date{21 December 2003}

\begin{abstract}
Counting statistics of charge transfers in a point contact interacting
with an arbitrary quantum system is studied. The theory for the charge
specific density matrix is developed, allowing the evaluation of the
probability of the outcome of any joint measurement of the state of
the quantum system and the transferred charge.  Applying the method of
charge projectors, the master equation for the charge specific density
matrix is derived in the tunneling Hamiltonian model of the point
contact.  As an example, the theory is applied to a quantum
measurement of a two-state system: The evolution of the charge
specific density matrix in the presence of Nyquist or Schottky noise
is studied and the conditions for the realization of a projective
measurement are established.  
\end{abstract}


\maketitle

\section{Introduction}\label{Intro}

In recent years, interest in quantum measurement has emerged in the
context of solid state devices.  This is in part forced by practical
issues in connection with experiments involving nanostructures, and
general questions of decoherence,\cite{BukSchHei98} and issues arising
in nanomechanics, for example in probing the quantum
electro-mechanical behavior of a nano-resonator, and is also crucial
for the read-out of a quantum computational process.  Solid state
nanodevices such as quantum dots are candidates for implementation of
spin qubits, \cite{LosDiV98,ImaAwsBur99,FujAusTok02,ElzHanGre98} and
charge qubits, \cite{HayFujChe03,DicLynJoh03} and superconducting
nanodevices containing Josephson junctions have been studied in
detail,\cite{MakSchSHn01} and are being tested for their potential as
charge qubits.  \cite{NakPasTsa99, VioAasCot02, DutGunBla03}
Nanomechanics and monitoring qubits thus confront us with the
practical details of a quantum measurement.\cite{WheZur83} In a
quantum measurement one must, for the system purported to function as
a detector, identify a collective variable which behaves classically.
Such a variable is not a priori provided by quantum theory, but in an
electronic device the candidate is ultimately related to the charge
flow.  It is therefore of importance to have a quantum description of the
charge transfer statistics, and demonstrate that a charge measurement
can be a measurement of the state of a quantum object coupled to it.
The purpose of the paper is to base such a description directly on the
density matrix of the many-body system.

The renewed interest in the quantum measurement process has in
particular focussed on a two-level system coupled to a quantum point
contact in the limit where the point contact can be modelled as a low
transparency tunnel junction.  The realization of the two-level system
could be two coherently coupled quantum dots between which an excess
electronic charge can tunnel.\cite{HayFujChe03} The two dots are
electrostatically coupled to the tunnel junction, or quantum point
contact, thereby making the tunnelling amplitude depend on the state
of the two-level system.  This model has recently attracted much
attention.
\cite{G97,GMWS01,GM01,K01,RK03,SMM02,BHO03}
Gurvitz
\cite{G97} derived for the zero temperature case, a Markovian master
equation for the density matrix of the two-level system keeping track
of the charge transferred through the junction.  Goan {\em et. al.}
\cite{GM01} considered the Bloch-Redfield equation for the spin
dynamics, i.e., the electron degrees of freedom of the tunnel junction
are traced out. To account for individual tunnelling events they
employed the quantum jump approach often used in quantum optics, and
obtained from the stochastic master equation the spin evolution for
specific realizations of tunnelling events.\cite{GMWS01} To account
for specific realizations of tunnelling events, Korotkov similarly
employed a stochastic treatment and showed by numerical simulation how
an initially mixed spin state can evolve into a pure state.\cite{K01}
Ruskov and Korotkov considered the Markovian master equation for the
density matrix for the two-level system keeping track of the charge
transferred through the junction at finite temperatures and calculated
the noise spectrum.\cite{RK03} Recent achievements in counting
statistics of charge transfers,\cite{LevLeeLes96} were used by
Shnirman {\em et. al.},\cite{SMM02} who derived a master equation for
the spin density matrix, keeping track of the charges passing through
the tunnel junction by a counting field as practised in counting
statistics, and calculated the noise spectrum of a quantum point
contact, with the same result as obtained by Bulaevskii {\em
et. al.}\cite{BHO03} Recently, a quantum oscillator coupled to the
junction has been studied, but by a method only valid at zero
temperature.\cite{MozMar02} The witnessed variety in techniques, calls
for a standard method to attack these types of problems.
        
We shall in the following develop a regular method for a joint
description of the charge kinetics of a tunnel junction and a quantum
system coupled to it.  Essential to the approach is that for an
arbitrary many-body system we show how to treat the number of
particles in a given spatial region or piece of material as a degree
of freedom.  This is achieved by employing suitably constructed charge
projection operators introduced previously in the context of counting
statistics.\cite{SheRam03} We shall refer to this reduced description,
where at any moment in time the probability distribution for the
number of particles or charges in a chosen region is specified, as the
charge representation.  The key construct of the theory is the charge
specific density matrix where the degrees of freedom of the
environment are partially traced out.  The charge specific density
matrix allows one to evaluate the probability of the outcome of any
joint measurement of the system and the charge state.  Applying the
charge projectors, allows us to use a standard kinetic approach to
obtain the equation of motion for the density matrix in the charge
representation.

Tunnel junctions and quantum point contacts are ubiquitous in
nanodevices such as for example those utilizing two-dimensional
electron gases in semiconductor inversion layers combined with split
gate technique.  The developed charge specific density matrix
description of the dynamics of a quantum object we believe is the
optimal description of such nanodevices since in electrical
measurements any information beyond the charge distribution is
irrelevant.  This sets the stage for considering questions regarding
quantum measurements, for example whether a measurement of the charge
can provide a measurement of the state of a quantum object coupled to
the junction.  Indeed we shall demonstrate that the junction is able
to perform a projective von Neumann measurement.  The model allows us
to study analytically a quantum measurement in the proper language of
the density matrix.  The intrinsic quantum bound on the measurement
time necessary for the tunnel junction to operate as a measuring
device is established.  Quantum theory and measurement has a long and
controversial history, and the model provides another simple
illustration that a quantum measurement can be described in standard
quantum mechanical terms, in fact in full detail for a realistic model
of a nanodevice.  The measuring scheme illustrates that there is no
need to postulate the classicallity of any variable, the charge state
of the junction being directly accessible, nor to invoke ``wave
function collapse.''  The transmission of the electrical noise in the
tunnel junction to the quantum object turns out to be sufficient for
the density matrix to decohere in the pointer basis.  Amplification
from the quantum to the classical level, the emergence of a projective
measurement, can thus be followed in a realistic model of a
nanodevice.

The paper is organized as follows: in section II we construct the
charge representation. In section III, the coupled object-junction
Hamiltonian is introduced, and in section IV the equation of motion in
the object and charge variables is obtained.  In section V, we briefly
discuss an isolated tunnel junction in the charge representation.  The
method we present is quite general, but for illustration we shall in
this paper apply it to a low transparency tunnel junction coupled to a
two-level system or for short a spin, and thereby obtain a master
equation in terms of the relevant variables, charge and spin. In
section VI, a short time measurement of the two-level system is
considered and an analytic solution of the master equation is
obtained.  Expressions for the characteristic times for decoherence
and spin-charge separation are obtained.  In section VII, the temporal
progression of a quantum measurement is studied and the emergence of a
projective measurement is seen to be different depending on the
relationship between voltage and temperature.  In section VIII, we
estimate the intrinsic quantum bounds on the measurement
time. Finally, we summarize and conclude. Details of calculations are
presented in appendices.

\section{Charge representation}\label{Charge representation}

We start by showing that the number of particles in a given spatial
region can be treated as a degree of freedom.  We construct the
probability distribution for the number of particles in a specified
region for an arbitrary many-body system.  This is accomplished by
using charge projection operators originally introduced in
Ref.~[\onlinecite{SheRam03}].

Consider an N-particle quantum system in a volume $V$ and in state
$\Psi $.  According to quantum mechanics, the probability,
$p_{n}^{A}$, that $n$ particles are in a sub-volume $A$ is
\begin{equation}
p_{n}^{A}= \int\limits_{V}
d\bm{r}_{1}
\ldots
d\bm{r}_{N} \;|
\Psi(\bm{r}_{1},\ldots,\bm{r}_{N})
|^2
\;  
\Theta_{n,N}^{A}(\bm{r}_{1},
\ldots,
\bm{r}_{N})
\label{4sd}
\end{equation}
where the indicator function $\Theta_{n,N}^{A}(\bm{r}_{1}, \ldots,
\bm{r}_{N})$ is equal to $1$ if exactly $n$ of its $N$ spatial
arguments belong to the region $A$, and is zero otherwise.  The volume
$A$ and its supplement volume $B$ partitions the total volume into two
non-overlapping parts, and the probability that there are $N-n$
particles in region $B$ equals the probability for having $n$
particles in region $A$.

The probability $p_{n}^{A}$ can be written as the expectation value
\begin{equation}
p_{n}^{A} = \langle \Psi | {\cal P}_{n}^{A}| \Psi \rangle \; ,
\label{5sd}
\end{equation}
of the Hermitian operator ${\cal P}_{n}^{A}$ which acts in the
position representation on the wave function according to
\begin{equation}
{\cal P}_{n}^{A}
\Psi(\bm{r}_{1},\ldots,\bm{r}_{N})
= 
\Theta_{n,N}^{A}(\bm{r}_{1},
\ldots,
\bm{r}_{N})
\Psi(\bm{r}_{1},\ldots,\bm{r}_{N}) .
\label{6sd}
\end{equation}
From their definition, one observes that the introduced operators,
${\cal P}_{n}^{A}$, have the properties
\begin{equation}
{\cal P}_{n}^{A}{\cal P}_{n'}^{A} = \delta_{nn'}{\cal P}_{n}^{A}
\quad,\quad \sum\limits_{n}{\cal P}_{n}^{A}=1
\label{8sd}
\end{equation}
and therefore constitute a complete set of Hermitian projectors.  The
projected state $|\Psi_{n}\rangle = {\cal P}_{n}^{A} |\Psi \rangle $
is the component of $ |\Psi \rangle $ for which exactly $n$ of the
particles are in region $A$.

The many-particle operator ${\cal P}_{n}^{A}$ can be expressed in a
simple way through single particle operators. We introduce the gauge
transformation operators
\begin{equation}
U_{\lambda}^{A} = \exp \left[ 
i \lambda \sum\limits_{k=1}^{N}\theta^{A}(\bm{r}_{k})
\right]
\; ,
\label{7sd}
\end{equation}
where $\lambda $ is a real parameter, and $\theta^{A}(\bm{r}_{k})$
equals $1$ if $\bm{r}_{k}$ belongs to the region $A$ and is zero
otherwise.  By virtue of the identity,
\begin{equation}
\int\limits_{0}^{2\pi} \frac{d \lambda }{2\pi }
e^{-i\lambda  n}\;U_{\lambda}^{A} = \;\Theta_{n,N}^{A}\,
\label{8sdd}
\end{equation}
the projection operator can be presented in any representation as
\begin{equation}
{\cal P}_{n}^{A} = \int\limits_{0}^{2\pi} \frac{d \lambda }{2\pi }
e^{- i n \lambda }U_{\lambda }^{A} \;,
\label{9sd}
\end{equation}
thus presenting the many-particle operator ${\cal P}_{n}^{A}$ in terms
of the product of the single particle gauge transformations in
$U_{\lambda}^{A}$.  This result can also be obtained by observing that
the projected state, $|\Psi_{n}\rangle $, is an eigenstate of
$U_{\lambda}^{A}$, $U_{\lambda}^{A} |\Psi_{n}\rangle= e^{i \lambda
n}|\Psi_{n}\rangle$.  If the region $A$, where the particles are
counted, occupies the whole volume $V$, the charge projector coincides
with the one used by P.W. Anderson in the theory of superconductivity
for projecting the BCS-wave function onto the state with a fixed
particle number.

\subsection{The charge specific density matrix}

We shall quite generally consider a system consisting of two parts: a
quantum object to be measured and the measuring device, which is taken
to be a tunnel junction connecting two electron reservoirs.

For the sake of the derivation, we assume that the object together
with the electron reservoirs constitute a closed quantum system. The
system is described by its full density matrix $\rho (\xi
,\bm{R}_{el}; \xi' ,\bm{R}_{el}')$, where $\xi $ is the coordinate of
the object and $\bm{R}_{el}= {\bm{r}_{1},\ldots \bm{r}_{N}}$ comprises
the electron coordinates, $N$ being the total number of electrons
relevant for the functioning of the measuring device.  With the
understanding that the only accessible information about the electrons
is their charge distribution, we introduce the charge specific density
matrix of the measured system
\begin{equation}
\hat{\rho }_{n}= {\rm{Tr}}_{el}\;( {\cal P}_{n}\; \rho )
\label{10sd}
\end{equation}
where ${\cal P}_{n}$ is the projection operator introduced in the
previous Section, and the trace is with respect to the electron
degrees of freedom $\bm{R}_{el}$.  The charge specific density matrix
$\hat{\rho }_{n} = \rho_{n}( \xi , \xi')$ allows one to deduce the
probabilities of any joint measurement performed on the quantum object
simultaneously with a charge measurement.

If the charge specific density matrix is traced over the remaining
degrees of freedom, the probability $p_{n}$ that there are $n$ charges
in a specified region (reference to which is suppressed in the
following) is the expectation value of the charge projector, or
expressed in terms of the charge specific density matrix
\begin{equation}
p_{n}= {\rm{Tr}_{\xi}}\;( \hat{\rho }_{n})
\label{jsd}
\end{equation}
where the trace is with respect to the degrees of freedom of the
quantum object, i.e., $\xi$.  It follows from Eq. (\ref{8sd}) that the
probability distribution is normalized, $\sum_{n}p_{n}=1$.  Apart from
proper normalization, $\hat{\rho }_{n}$ is the density matrix for the
quantum object after a charge measurement.

The charge projection operators define what we shall refer to as the
charge representation.  The degree of freedom in the charge
representation is the variable $n$: the number of charges in a
specified spatial region or equivalently in a piece of material. All
other information about the charges is traced out.  The charge
representation is a strongly reduced representation, the description
of the environment is reduced to one variable, but if interest is in
the currents in a system, information beyond the charge representation
is irrelevant.

We now consider the charge specific dynamics, i.e., the equation of
motion of the quantum object for specified charge.  The dynamics of
the combined system is governed by the equation for the density matrix
\begin{equation}
i \dot{\rho }= [H, \rho ]
\label{x4c}
\end{equation}
and the equation of motion for the charge specific density matrix is
thus
\begin{equation}
i \dot{\hat{\rho } }_{n}= {\rm{Tr}}_{el}\;
\left(
{\cal P}_{n} [H, \rho ]
 \right) .
\label{y4c}
\end{equation}
The charge specific density matrix describes the quantum dynamics of
the coupled object conditioned on the charge variable being specified.
Since the charge specific description is a reduced description,
generally a hierarchy of equations is generated.  In the following we
shall study a situation where ${\hat{\rho } }_{n}$ can be obtained
explicitly.

\section{Model Hamiltonian }\label{ham}

In this section we discuss the Hamiltonian for an arbitrary quantum
system coupled to a tunnel junction
\begin{equation}
\hat{H} = 
\hat{H}_{S}+
H_{l}+
H_{r}+
\hat{H}_{T}
.
\label{t4c}
\end{equation}
Here, $\hat{H}_{S}$ is the Hamiltonian for the isolated quantum
 object, the system to be measured,
 and a hat
marks terms which are operators 
with respect to the degrees of freedom 
of the quantum object.  The Hamiltonians
$H_{l,r}$ specify the isolated left and right electrodes of the
junction
\begin{equation}
H_{l} = \sum\limits_{\bm{l}}\varepsilon_{\bm{l}}
c_{\bm{l}}^{\dagger} c_{\bm{l}}
\quad,\quad 
H_{r} = 
\sum\limits_{\bm{r}}\varepsilon_{\bm{r}} c_{\bm{r}}^{\dagger} c_{\bm{r}}
\label{v4c}
\end{equation}
where $\bm{l,r}$ labels the quantum numbers of the single particle
energy eigenstates in the left and right electrodes, respectively,
with corresponding energies $\varepsilon_{\bm{l,r}}$ and annihilation
and creation operators.  The operator $\hat{H}_{T}$ describes
tunnelling,
\begin{equation}
\hat{H}_{T}=
\hat{{\cal T}}
+
\hat{{\cal T}}^{\dagger}
\label{cdd}
\end{equation}
where
\begin{equation}
\hat{{\cal T}}
=
\sum\limits_{\bm{l}, \bm{r}} 
\hat{T}_{\bm{l}\bm{r}} 
c_{\bm{l}}^{\dagger}c_{\bm{r}}
\quad,\quad 
\hat{{\cal T}}^{\dagger}
=
\sum\limits_{\bm{l}, \bm{r}} 
\hat{T}_{\bm{r}\bm{l}} 
c_{\bm{r}}^{\dagger}c_{\bm{l}}
\label{edd}
\end{equation}
and $\hat{T}_{\bm{l}\bm{r}}$ are operators acting on the coordinates,
$\xi $, of the quantum object.  The Hermitian property of the
Hamiltonian requires that $ \hat{T}_{\bm{l}\bm{r}} =
\hat{T}_{\bm{r}\bm{l}}^{\dagger} $.  Compared to an isolated tunnel
junction, the additional feature of the model is that the tunnelling
amplitude, and thereby the conductance of the tunnel junction, depends
on the state of the measured system.

In the next section we shall study the dynamics of the system when the
probability distribution for the number of charge transfers through
the junction is specified at all times.  This is achieved by using the
charge projectors introduced in section \ref{Charge representation}.
For a tunnel junction, the spatial region of interest for counting
charges is either of the two electrodes, say we choose the left one.
Of importance are in view of Eq.~(\ref{y4c}) the commutation relations
of the charge projectors and the Hamiltonian. The terms in the
Hamiltonian commute with the charge projectors except for the
tunnelling term.  The discrete charge dynamics of the tunnel junction
is specified by the charge projection operators according to
\begin{equation}
{\cal P}_{n} c_{\bm{l}}^{\dagger}c_{\bm{r}} =
 c_{\bm{l}}^{\dagger}c_{\bm{r}}{\cal P}_{n-1}
\quad,\quad 
{\cal P}_{n} c_{\bm{r}}^{\dagger}c_{\bm{l}} =
 c_{\bm{r}}^{\dagger}c_{\bm{l}}{\cal P}_{n+1} .
\label{jdd}
\end{equation}
In terms of the tunnelling operators, Eq.~(\ref{edd}), the
identities read
\begin{equation}
{\cal P}_{n} \hat{{\cal T}}
=
\hat{{\cal T}}{\cal P}_{n-1}
\quad,\quad 
{\cal P}_{n} \hat{{\cal T}}^{\dagger}
=
\hat{{\cal T}}^{\dagger}{\cal P}_{n+1} .
\label{ndd}
\end{equation}
These identities are used repeatedly in the derivation of the equation
of motion in the charge representation.

\section{Charge specific dynamics}\label{spichar}

In this section we shall obtain the equation of motion for the charge
specific density matrix, $\hat{\rho }_{n}$, to lowest order in the
tunnelling.

Taking advantage of the relations in Eq.~(\ref{jdd}), the equation of
motion for the charge specific density matrix, Eq.~(\ref{y4c}), can be
written in the form
\begin{eqnarray}
 \dot{\hat{\rho }}_{n}(t) &+& i[ \hat{H}_{S},\hat{\rho }_{n}(t) ]
\nonumber \\&=&  
\sum\limits_{\bm{l}\bm{r}}
\left(
\hat{T}_{\bm{l}\bm{r}}
\hat{A}_{\bm{l}\bm{r}}^{(n)}(t)
+ 
\hat{T}_{\bm{r}\bm{l}}
\hat{B}_{\bm{r}\bm{l}}^{(n)}(t)
\right) + h.c. 
\label{wddn}
\end{eqnarray}
where here and in the following $h.c.$ represents the hermitian
conjugate term, 
and the time
dependent system operators $\hat{A}$ and $\hat{B}$ are given by
\begin{eqnarray}
\hat{A}_{\bm{lr}}^{(n)}(t)= 
\frac{1}{i}
{\rm{Tr}}_{el}\;
\left(
 c_{\bm{l}}^{\dagger}c_{\bm{r}}\rho(t){\cal P}_{n}
 \right),\\
\hat{B}_{\bm{lr}}^{(n)}(t) = 
\frac{1}{i}
{\rm{Tr}}_{el}\;
\left(
 c_{\bm{r}}^{\dagger}c_{\bm{l}}\rho(t){\cal P}_{n}
 \right) .
\label{xddd}
\end{eqnarray}
One sees from Eq.~(\ref{wddn}), that the time evolution of the charge
diagonal component of the density matrix, $ \hat{\rho }_{n} $, is
determined (in addition to the internal dynamics of the system) by the
charge off-diagonal components ${\cal P}_{n\pm 1}\rho {\cal P}_{}$
that control the matrices $\hat{A}$ and $\hat{B}$.  The latter two are
small, being generated by the rare tunnelling events, and can be
expressed in terms of the diagonal elements using perturbation theory.
To obtain, e.g. $\hat{A}$, one uses the equation of motion,
Eq.~(\ref{x4c}), to obtain
\begin{eqnarray}
 \dot{\hat{A}}_{\bm{lr}}^{(n)}(t)& -&
 i \omega_{\bm{lr}} \hat{A}_{\bm{lr}}^{(n)}(t)
 + i [ \hat{H}_{S}, \hat{A}_{\bm{lr}}^{(n)}(t)] =\nonumber \\&-&
{\rm{Tr}}_{el}\;
\left(
 c_{\bm{l}}^{\dagger}c_{\bm{r}}[H_{T},\rho(t)]{\cal P}_{n}
 \right)
\label{4ddn}
\end{eqnarray}
where $\omega_{\bm{lr}}\equiv \varepsilon_{\bm{l}}-
\varepsilon_{\bm{r}}$. To lowest order in tunnelling, one evaluates
the source term on the r.h.s. retaining only the charge diagonal
${\cal P}_{m}\rho {\cal P}_{m}$ components ($m=n$ and $m=n -1$) of the
full density matrix $\rho $; the off-diagonal components, ${\cal
P}_{n-2}\rho{\cal P}_{n}$ and ${\cal P}_{n-1}\rho{\cal P}_{n+1}$, give
higher order corrections which are neglected (for details we refer to
appendix \ref{derv}).  The diagonal terms can be expressed in terms of
the single particle distribution functions for the electrodes,
$f_{\bm{l},\bm{r}}$, which, since the electrodes act as particle
reservoirs, can be taken independent of the total charge number of the
particles in the left electrode, $n$.  As shown in appendix
\ref{derv}, to lowest order in the tunnelling the inhomogeneous term
on the right in Eq.~(\ref{4ddn}) can be expressed in terms of the
charge specific density matrix and an explicit solution for $\hat{A}$
(and similarly for $\hat{B}$) can be obtained. The resulting equation
is a non-Markovian master equation for the charge specific density
matrix for the system, Eq.~(\ref{2ed}).  In the limit where the part
of the tunnelling matrix element in Eq.~(\ref{w4cx}) which depends on
the coupling to the system is small, the temporal non-locality of the
kernels can be neglected and the Markovian equation is obtained.  The
master equation for the charge specific density matrix
$\hat{\rho}_{n}$ has the form:
\begin{eqnarray}
 \dot{\hat{\rho }}_{n}(t)
   &=&  
\frac{1}{i} [(\hat{H}_{S} + \hat{{\cal M}}), \hat{\rho}_{n}(t)] \nonumber \\ 
&+& {\cal L}\sss{\hat{\rho}_{n}(t)}
+
{\cal D}\sss{\hat{\rho}_{n}''(t)}
+
{\cal J}\sss{\hat{\rho}_{n}'(t)}
\label{hjd}
\end{eqnarray}
where $\hat{\rho}_{n}'$ and $\hat{\rho}_{n}''$ denote ``discrete
derivatives'',
\begin{eqnarray}
\hat{\rho}_{n}' = \frac{1}{2}\left(
\hat{\rho}_{n+1} - \hat{\rho}_{n-1}\right)
,\\
\hat{\rho}_{n}'' = 
\hat{\rho}_{n+1}+ \hat{\rho}_{n-1} - 2 \hat{\rho}_{n}
.
\label{gjd}
\end{eqnarray}
The Hamiltonian $\hat{H}_{S}$ describes the dynamics of the isolated
system, and the rest of the terms on the r.h.s. of Eq.~(\ref{hjd})
describe the effect of the tunnelling events; in the formulae below,
the subscript $+(-)$ marks the contribution of the processes when an
electron tunnels from the left (right) electrode to the right (left)
one.  The ``magnetization'' term, $ \hat{{\cal M}} = \hat{{\cal
M}}_{+} + \hat{{\cal M}}_{-} $,
\begin{eqnarray}
\hat{{\cal M}}_{+}& =  &  
 \frac{1}{2i}\sum\limits_{\bm{l}\bm{r}}
f_{\bm{l}}(1- f_{\bm{r}})
\left(
\hat{T}_{\bm{l}\bm{r}}
\ssss{\hat{T}_{\bm{lr}}^{\dagger}} - h.c. \right)
            \label{ihd}\\   
\hat{{\cal M}}_{-}     & =  &  
\frac{1}{2i}\sum\limits_{\bm{l}\bm{r}}
 f_{\bm{r}}(1- f_{\bm{l}})
\left(
\hat{T}_{\bm{rl}}
\ssss{\hat{T}_{\bm{rl}}^{\dagger}} 
 - h.c.
\right) 
            \label{ihd2}
\end{eqnarray}
renormalizes the system Hamiltonian $\hat{H}_{S}$.  The operator
${\cal L}= {\cal L}_{+}+{\cal L}_{-}$, where ${\cal
L}_{\pm}\sss{\hat{\rho} }$ maps $\hat{\rho}$ according to
\begin{widetext}
\begin{equation}
{\cal L}_{+}\sss{\hat{\rho} }=
\sum\limits_{\bm{l}\bm{r}}
 f_{\bm{l}}(1- f_{\bm{r}})
\left(
 \ssss{
\hat{T}_{\bm{lr}}^{\dagger}}
\hat{\rho }
\hat{T}_{\bm{lr}}
  - \frac{1}{2}
\left[
\left(
\hat{T}_{\bm{lr}}
\ssss{
\hat{T}_{\bm{lr}}^{\dagger}
}
 \right)
,
\hat{\rho}
\right]_{+}
 + h.c.\right)
\label{6ud}
\end{equation}
and
\begin{equation}
{\cal L}_{-}\sss{\hat{\rho} }=
\sum\limits_{\bm{l}\bm{r}}
 f_{\bm{r}}(1- f_{\bm{l}})
\left(
\hat{T}_{\bm{lr}}
\hat{\rho }
\ssss{
\hat{T}_{\bm{lr}}^{\dagger}
} 
-\frac{1}{2}
\left[
\left(
\ssss{
\hat{T}_{\bm{lr}}^{\dagger}
}
\hat{T}_{\bm{lr}}
 \right)
,
\hat{\rho}
\right]_{+}
+ h.c. 
\right)\;.
\label{7ud}
\end{equation}
\end{widetext}
The diffusion, ${\cal D}$, and drift, ${\cal J}$, operators are
\begin{eqnarray}
{\cal D}\sss{\hat{\rho}} = 
\frac{1}{2}
\left(
{\cal
D}_{+}\sss{\hat{\rho}} + {\cal D}_{-}\sss{\hat{\rho}}
\right)
 \nonumber, \\
{\cal J}\sss{\hat{\rho}} = {\cal D}_{+}\sss{\hat{\rho}} - {\cal
D}_{-}\sss{\hat{\rho}}, 
\end{eqnarray} 
where
\begin{equation}
{\cal D}_{+}\sss{\hat{\rho} }=
\sum\limits_{\bm{l}\bm{r}}
 f_{\bm{l}}(1- f_{\bm{r}})
 \ssss{
\hat{T}_{\bm{lr}}^{\dagger}
}
\hat{\rho }
\hat{T}_{\bm{lr}}
 + h.c.
\label{xyedd}
\end{equation}
and
\begin{equation}
{\cal D}_{-}\sss{\hat{\rho} }=
\sum\limits_{\bm{l}\bm{r}}
 f_{\bm{r}}(1- f_{\bm{l}})
\hat{T}_{\bm{lr}} 
\hat{\rho }
\ssss{
\hat{T}_{\bm{lr}}^{\dagger}
} + h.c. 
\label{xzedd}
\end{equation}
We have introduced the notation (but suppressed the time dependence in
the above formulas)
\begin{equation}
\ssss{
\hat{T}_{\bm{lr}}^{\dagger}
}(t)
=
\int\limits_{0}^{t}d \tau 
\;
 e^{i
 \omega_{\bm{lr}}\tau +i\int\limits_{t- \tau }^{t} dt' V(t')  }
e^{-i \hat{H}_{S}\tau }
\hat{T}_{\bm{lr}}^{\dagger}
e^{ i\hat{H}_{S}\tau }
\label{rgdd}
\end{equation}
and assumed a voltage $U$ applied to the junction, $V(t) = e U(t)$,
$e$ being the electron charge.

The master equation, Eq.~(\ref{hjd}), requires an initial condition.
We assume that the left and right electrodes of the junction are
disconnected at times preceding the initial moment $t=0$, so that the
electrodes are in definite charge states, say with $n^{(0)}$ electrons
in the left electrode.  Using the convention that $n$ is counted from
$n^{(0)}$, the variable $n$ becomes the number of electrons {\em
transferred} from the right to the left electrode.  The initial
condition reads
\begin{equation}
\hat{\rho }_{n}(t=0) = \delta_{n,0}\; \hat{\rho}^{(0)}
\label{4ud}
\end{equation}
where $\hat{\rho}^{(0)}$ is the initial density matrix of the measured system.

The kernel ${\cal D}$ describes diffusion in charge space, and the
average charge current from the left to the right electrode, $- e
\sum\limits_{n} n\, {\rm{Tr}}\;\dot{\hat{\rho}}_{n}(t)$, is given by
the drift term ${\cal J}$
\begin{equation}
I(t) 
=
e{\rm{Tr}}\;{\cal J}\sss{\hat{\rho}(t)}
\label{pmd}
\end{equation}
where $\hat{\rho}$ is the (unconditional) density matrix
\begin{equation}
\hat{\rho} =\sum\limits_{n}\hat{\rho}_{n}(t) .
\label{qmd}
\end{equation}

We stress that the form of the derived master equation for the charge
specific density matrix is valid for any quantum object coupled to the
junction.  In the present paper, we shall consider the electrons
coupled to a two-level system or for short a spin, a system considered
in numerous papers as discussed in the
introduction,\cite{G97,GMWS01,GM01,K01,RK03,SMM02,BHO03} and
$\hat{\rho }_{n}$ will be a $2\times2$ matrix.  A remaining task is
thus to unravel the complex spin structure of the master equation
presently hidden in the bracket operation.  However, first we consider
the isolated tunnel junction.

\section{Isolated tunnel junction}\label{nospin}

To familiarize with the charge representation we pause to consider the
case where the system is decoupled from the junction.  In this case,
the only degree of freedom of interest is the charge $n$, and the
equation for the probability, $p_{n}$, for the transfer of $n$ charges
\begin{equation}
\dot{p}_{n} =  D p_{n}'' + I p_{n}'
\label{fqd}
\end{equation}
is obtained by 
taking trace of the
master equation Eq.~(\ref{hjd}). Here,
\begin{equation}
I =
2\pi  \sum\limits_{\bm{l}\bm{r}}
|T_{\bm{lr}}|^2
( f_{\bm{l}}- f_{\bm{r}})
\delta (\varepsilon_{\bm{l}} + V -\varepsilon_{\bm{r}}) .
\label{dfd}
\end{equation}
and
\begin{equation}
 D =
 \pi \sum\limits_{\bm{l}\bm{r}}
|T_{\bm{lr}}|^2
( f_{\bm{l}}+f_{\bm{r}}
- 2 f_{\bm{l}}f_{\bm{r}}
)
\delta (\varepsilon_{\bm{l}} + V -\varepsilon_{\bm{r}})
.
\label{iqd}
\end{equation}
For simplicity, we assume the bias $V$ is time independent.

The physical meaning of the parameters $I$ and $D$ in Eq.~(\ref{fqd})
can be deduced from the structure of the equation.  Multiplying
Eq.~(\ref{fqd}) by $n$ and performing summation with respect to $n$,
one readily sees that $I$ equals the {\em dc} current through the
junction, i.e., the average rate of charge transfers from the left to
the right electrode, $I = - \frac{d }{dt}\sum\limits_{n}n p_{n}(t)$.
The value of the {\em dc} current $I= I(V)$ is given by
Eq.~(\ref{dfd}).  For the time evolution of the variance of the charge
distribution,
\begin{equation}
\Delta n(t) = \langle n^2(t)\rangle - \langle n(t)\rangle^2 \; ,
\label{r3d}
\end{equation}
where $\langle n^k(t)\rangle = 
 \sum\limits_{n}n^k p_{n}(t)
$, one obtains from Eq.~(\ref{fqd})
\begin{equation}
\frac{d \Delta n }{dt} = 2D \;,
\label{s3d}
\end{equation}
thus $D$ has the meaning of the charge diffusion coefficient,
characterizing the randomness of the charge transfers.

To connect with the standard noise discussion in terms of current
fluctuations, we express the transferred charge, $n(t)$ as the time
integral of the current $i(t)$, $ n(t)= \int_{0}^{t} dt' i (t')$, and
the variance takes the form
\begin{equation}
\Delta n(t) =
2\int\limits_{0}^{t} dt' \int\limits_{0}^{t'}  d \tau \,
S(\tau )
\label{cgd} 
\end{equation}
where $S(\tau )= S(- \tau )$ is 
 the current-current correlator, 
\begin{equation}
S(\tau ) = \langle
\delta i(t)\delta i(t+ \tau ) \rangle
\quad,\quad 
\delta i(t) = i(t) - \langle i \rangle .
\label{t3d}
\end{equation}
At times much larger than the current-current correlation time, one
obtains from  Eq.~(\ref{cgd})
\begin{equation}
\frac{d \Delta n}{dt} = 2\int\limits_{0}^{\infty }d \tau \, S(\tau ) \;.
\label{u3d}
\end{equation}
Recalling Eq.~(\ref{s3d}), one establishes the relation,
\begin{equation}
4D = S_{\omega =0} \;,
\label{v3d}
\end{equation}
between the charge diffusion coefficient $D$ and the power spectrum of
the current noise, $S_{\omega } = 2 \int_{-\infty }^{\infty }d \tau
S(\tau) \cos(\omega \tau)$, at zero frequency.

In quasi-equilibrium, when $f_{\bm{l,r}}$ are Fermi functions
corresponding to the temperature $T$, one readily obtains from
Eqs.~(\ref{dfd}) and (\ref{iqd}) that the charge diffusion constant
$D$ is related to the {\em dc} current $I$ as $2D = I \coth V/2T$.  In
view of Eq.~(\ref{v3d}), this corresponds to the well-known
result,\cite{Kog96}
\begin{equation}
S_{\omega =0} = 
2I(V) \coth \frac{V}{2T}
\label{tmd}
\end{equation}
which expresses the fluctuation-dissipation theorem (and holds for
arbitrary voltages for the case of a tunnel junction ).\cite{nonmark}

Given the initial charge distribution, Eq.~(\ref{fqd}) can be solved
introducing the Fourier transform $\chi (\lambda ;t)=
\sum_{n}p_{n}(t)e^{i \lambda n}$.  For the initial condition
$p_{n}(t=0)= \delta_{n,0}$, the solution reads
\begin{equation}
\chi (\lambda ;t) = \exp[2 D t(\cos \lambda -1) - i I t \sin \lambda ] \;.
\label{53d}
\end{equation}
Not surprisingly, this expression reproduces the long time (Markovian)
limit \cite{full} of the generating function of the full counting
statistics for a tunneling junction first derived in
Ref.~\onlinecite{LevRez01}.  Inverting the Fourier transform (see
appendix \ref{solv2} for details), the probability of $n$-charge
transfers in the time span $t$ reads
\begin{equation}
p_{n}(t) = I_{n}\left(\frac{\langle n \rangle}{\sinh v} \right)
\exp\left[v n - \langle n \rangle \coth v\right]
\label{63d}
\end{equation}
where $I_{n}$ is the modified Bessel function, $\langle n \rangle =
I(V)t$ is the average charge transfer for the bias $V$, and $v =
V/2T$.  The moments of the distribution, $\langle n^{k}\rangle$ and
$(n - \langle n \rangle)^{k}$, can be evaluated with the help of the
generating functions Eq.~(\ref{8pd}) in appendix \ref{solv2}.

\section{Short time measurement}\label{short}

The charge specific density matrix can be used to describe a quantum
measurement in a two-fold way: (i) measuring the charge transferred
through the junction provides information about the state of the
system coupled to the junction, or (ii) a measurement of the coupled
system can reveal information about the charge state of the junction.
In this paper we shall consider the case where the system coupled to
the junction is a two-level system or for short a spin, whose internal
dynamics is specified by $\hat{H}_{S} = \Omega \sigma_x$.  We are
interested in the read-out of the state of the spin, and we analyze
how measuring the charge can be a projective measurement of the spin.

For definiteness, we chose to measure the $z$-projection of the spin,
and for the measurement in this basis the coupling to the spin is
taken to be
\begin{equation}
\hat{T}_{\bm{l}\bm{r}}=
v_{\bm{l}\bm{r}} +
\hat{\sigma}_{z}w_{\bm{l}\bm{r}}   
\label{w4cx}
\end{equation}
so that the spin up and down states correspond to extreme values in
tunnelling strength. It is convenient to introduce the following
quantities:
\begin{equation}
\left\{
\begin{array}{c}
   G_{V}\\
   G_{W}\\
   G_{1}\\
   G_{2}
 \end{array}
\right\}
=
2\pi  \sum\limits_{\bm{l}\bm{r}}
\left\{
\begin{array}{c}
V_{\bm{lr}}\\
W_{\bm{lr}}\\
U_{1\bm{lr}}\\
U_{2\bm{lr}}\\
 \end{array}
\right\}
\left(
-\frac{\partial f(\varepsilon_{\bm{l}})}{\partial \varepsilon_{\bm{l}}}
 \right)
\delta (\varepsilon_{\bm{l}}  -\varepsilon_{\bm{r}})\;,
\label{4jd}
\end{equation}
where
$V_{\bm{lr}}= |v_{\bm{lr}}|^2$, $W_{\bm{lr}}= |w_{\bm{lr}}|^2$, 
$U_{1\bm{lr}}= \Re v_{\bm{lr}}^{*}w_{\bm{lr}}$, 
and $U_{2\bm{lr}}= \Im v_{\bm{lr}}^{*}w_{\bm{lr}}$. 
For the up and down spin orientation, the tunneling conductance 
(in the units $e^2/ \hbar $) is
given by $G_{V}+ G_{W} \pm 2 G_{1}$, respectively.
 
Following von Neumann,\cite{Neu32} we aim at an effectively
instantaneous measurement of the spin quantum state, which means that
the time of the measurement must be short compared to the intrinsic
spin precession time.  To describe the measurement process, we are
therefore interested in solving the master equation for times much
shorter than the inverse of the Rabi frequency $\Omega $.  In that
case, $\hat{H}_{S}$ in Eq.~(\ref{hjd}) and Eq.~(\ref{rgdd}) can be set
to zero, and the spin structure of the master equation,
Eq.~(\ref{hjd}), reduces considerably.

In this limit, the renormalization term $\hat{\cal M}= \hat{\cal
M}_{+} + \hat{\cal M}_{-}$, defined by Eqs.~(\ref{ihd}), and
(\ref{ihd2}) is $\hat{{\cal M}}= M_{z}\hat{\sigma}_{z}$ where
\begin{equation}
M_{z} = 
2 \sum\limits_{\bm{l}\bm{r}}
\Re v_{\bm{lr}}^{*}w_{\bm{l}\bm{r}}
 \frac{ f_{\bm{l}} - f_{\bm{r}}}
{\varepsilon_{\bm{l}} + V -\varepsilon_{\bm{r}}} .
\label{w3d}
\end{equation}
The physical mechanism behind the renormalization of the spin
``Zeeman'' energy given by $M_{z}$ is the sensitivity of the total
{\em electron} energy to the spin orientation. Indeed, in second order
perturbation theory with respect to tunneling, the single electron
levels acquire a shift controlled by the square modulus of the
tunneling matrix element which is either $|v_{\bm{lr}} +
w_{\bm{lr}}|^2$, when the spin is up, or $|v_{\bm{lr}} -
w_{\bm{lr}}|^2$, when the spin is down. The interference term 2$\Re
v_{\bm{lr}}^{*} w_{\bm{l}\bm{r}}$ gives rise to the spin dependence of
the energy and, therefore, $M_{z}$ in Eq.~(\ref{w3d}).  The sign of
the combination $\Re v_{\bm{lr}}^{*} w_{\bm{l}\bm{r}}$ is not fixed by
any physical requirement.  In a mesoscopic system, we expect it to be
a random function of the quantum numbers $\bm{l}$ and $\bm{r}$,
fluctuating both in absolute value and sign.  In view of this, the
contributions to $M_{z}$ from the states in the wide energy range of
the order of the Fermi energy, have the tendency for cancellation. As
an estimate, $M_{z} \sim G_{1} \delta E $ where $\delta E$ is a narrow
energy interval within which $\Re v_{\bm{lr}}^{*} w_{\bm{l}\bm{r}}$
are correlated.  The renormalization may therefore be a small
correction, which we neglect below.

The master equation, Eq.~(\ref{hjd}), becomes
\begin{eqnarray}
\dot{\hat{\rho}}_{n}& =   &  \frac{G_2 V}{i}\left[\hat{\sigma}_{z},
     \hat{\rho}_{n}\right]    
+2G_{W} V \coth\frac{V}{2T}\left(
\hat{\sigma}_{z}\hat{\rho}_{n}\hat{\sigma}_{z} - \hat{\rho}_{n}
 \right)
          \nonumber \\   
     & +  &
\frac{1}{2}V \coth\frac{V}{2T}
\left(
G_{V}\hat{\rho}_{n}''+ 
G_{1}[\hat{\rho}_{n}'', \hat{\sigma}_{z}]_{+} 
+ G_{W} \hat{\sigma}_{z}\hat{\rho}_{n}''\hat{\sigma}_{z}
 \right) \nonumber \\
& + & 
V 
\left(
G_{V}\hat{\rho}_{n}'+ 
G_{1}[\hat{\rho}_{n}', \hat{\sigma}_{z}]_{+} 
+ G_{W} \hat{\sigma}_{z}\hat{\rho}_{n}'\hat{\sigma}_{z}
 \right) \nonumber \\
&+&i G_{2} V[\hat{\rho}_{n}'', \hat{\sigma}_{z}]
+ i G_{2} V \coth\frac{V}{2T} [\hat{\rho}_{n}', \hat{\sigma}_{z}],              
\label{qkd}  
\end{eqnarray}
where 
$G_{V,W,1,2}$  are defined in Eq.~(\ref{4jd}).

At short times, the components of the density matrix,
\begin{equation}
\hat{\rho}_{n}= \left(
\begin{array}{lr}
u_{n}&   \alpha_{n}\\
 \alpha_{n}^{*}& d_{n}
\end{array}
\right) \;,
\label{skd}
\end{equation}
are decoupled, and solving Eq.~(\ref{qkd}) amounts to solving the
following three equations:
\begin{eqnarray}
\dot{u}_{n} &=& 
\frac{1}{2}V \coth\frac{V}{2T} (G_{V}+ G_{W} + 2 G_{1})u_{n}''\nonumber \\
&+&
V  (G_{V}+ G_{W} + 2 G_{1})u_{n}'\;,
\label{tkd}
\end{eqnarray}
\begin{eqnarray}
\dot{d}_{n} &=& 
\frac{1}{2}V \coth\frac{V}{2T} (G_{V}+ G_{W} - 2 G_{1})d_{n}''\nonumber \\
&+&
V  (G_{V}+ G_{W} - 2 G_{1})d_{n}'\;,
\label{vkd}
\end{eqnarray}

\begin{eqnarray}
\dot{\alpha}_{n}& =& - 2 i G_2 V \alpha_{n}- 2 G_{W} V \coth\frac{V}{2T}\; \alpha_{n}
              \nonumber \\   
     & +  &
\frac{1}{2}\left( 
V \coth\frac{V}{2T} (G_{V}- G_{W}) - 2 i  G_{2}V\right)\alpha_{n}''
             \nonumber \\   
     &  + & 
\left(
V  (G_{V}- G_{W}) - 2 i G_{2}V \coth\frac{V}{2T} \right) \alpha_{n}'
 .
            \label{wkd}
\end{eqnarray}

Solving the master equation, Eq. (\ref{qkd}), is thus achieved once
the equation with the structure
\begin{equation}
\dot{x}_{n} =  D_{x}x_{n}'' +  J_{x} x_{n}' 
\label{xkdd}
\end{equation}
is solved.  According to appendix \ref{onecomp}, the solution is
specified in terms of the modified Bessel function
\begin{equation}
x_{n}(t) = x_{0}
\left(\frac{D_{x-}}{D_{x+}} \right)^{n/2}\
e^{- 2 D_{x} t }I_{n}(2t \sqrt{D_{x+}D_{x-}})
\label{bpd}
\end{equation}
where $D_{x\pm}= D_{x}\pm \frac{1}{2} J_{x}$.
All moments
of the stochastic process $x_{n}(t)$ can be expressed in terms of the
parameters $D_{x}$ and $J_{x}$ which for the original problem becomes
in terms of voltage, temperature and time.

\subsection{Characteristic times}\label{}

Let us now analyze the time dependence of the charge specific density
matrix. It is found solving the master equation with the initial
condition Eq.~(\ref{4ud}), where $\hat{\rho}^{(0)}$ is the initial spin
state (to be measured),
\begin{equation}
\hat{\rho}^{(0)} = 
\left(
\begin{array}{lr}
u_{0}&  \alpha_{0}\\
\alpha_{0}^{*}& d_{0}
\end{array}
\right) .
\label{9ud}
\end{equation}

First, we consider the unconditional spin density matrix $\rho $,
Eq.~(\ref{qmd}), which gives information about the spin state
irrespective of the outcome of the charge measurement ({\it i.e.},
discarding the results of a charge measurement).  The equation for
$\hat{\rho}$,
\begin{equation}
\dot{\hat{\rho}} =     \frac{G_2 V}{i}\left[\hat{\sigma}_{z},
     \hat{\rho}\right]
+2G_{W} V \coth\frac{V}{2T}\left(
\hat{\sigma}_{z}\hat{\rho}\hat{\sigma}_{z} - \hat{\rho}
 \right)\;,
\label{8ud}
\end{equation}
is obtained by summation with respect to $n$ in Eq.~(\ref{qkd}). Its
solution reads,
\begin{equation}
\hat{\rho}(t) = 
\left(
\begin{array}{lr}
u_{0}&  \alpha_{0}e^{-2i G_2 Vt}e^{- t/\tau_{d}}\\
\alpha_{0}^{*}e^{2i G_2 Vt}e^{- t/\tau_{d}}& d_{0}
\end{array}
\right)
\label{avd}
\end{equation}
where the parameter $\tau_{d}$,
\begin{equation}
\tau_{d}^{-1}= 
2G_{W} V \coth\frac{V}{2T}\; ,
\label{bvd}
\end{equation}
is the decoherence time, which gives the decay rate of the charge
unconditional off-diagonal elements of the density matrix
$\sum_{n}\alpha_{n}$.

As time passes, the probability distributions for being in the spin up
or down states, $u_{n}(t)$ and $d_{n}(t)$, will according to
Eq. (\ref{bpd}) drift and broaden in charge space.  To investigate the
different drift in the two distributions we consider their overlap in
charge space, $\sum\limits_{n}u_{n}(t)d_{n}(t)$ and we get according
to Eq.~(\ref{dvd})
\begin{equation}
\sum\limits_{n}u_{n}(t)d_{n}(t)=
u_{0}d_{0}
e^{- 2 Vt
\coth\frac{V}{2T}(G_{V}+ G_{W})}I_{0}
(\tilde{t})
\label{wnd}
\end{equation}
where
$
\tilde{t}=
2V t\coth\frac{V}{2T} 
\sqrt{(G_{V}+ G_{W})^2 - 4 G_{1}^2
\tanh^2\frac{V}{2T}} .
$
At large times, $\tilde{t}\gg 1$, the overlap decays exponentially,
$e^{- t/ \tau_{z}}$, with the characteristic rate
\begin{eqnarray}
\tau_{z}^{-1}&=&  
8 V \tanh\frac{V}{2T}\; \frac{G_{1}^2}{G_{V}+ G_{W}} \nonumber \\ & \times & \left(1 +
 \sqrt{1 -\frac{4 G_{1}^2 \tanh^2\frac{V}{2T}}{(G_{V}+ G_{W})^2} }
\right)^{-1}\;.
\label{frd}
\end{eqnarray}
After the time span $\tau_{z}$, the probability distributions for spin
up and down have separated in charge space.\cite{vanishG1}

A similar calculation for the off-diagonal elements of the charge
specific spin density matrix gives
\begin{equation}
\sum\limits_{n}|\alpha_{n}(t)|^2 =
|\alpha_{n}(0)|^2e^{-2t (G_{V}+G_{W}) V \coth\frac{V}{2T}}\; 
I_{0}(t^{*})\,,
\label{bnd}
\end{equation}
where $t^{*}=
2t V \coth\frac{V}{2T}\left|G_{V}-G_{W} + 2i G_{2}\right|$,
and at large times the off-diagonal elements have decayed
exponentially at the characteristic decoherence rate
\begin{equation}
\tau_{\perp}^{-1}=
2V \coth\frac{V}{2T}
\left(
G_{V}+G_{W}
- \left|G_{V}-G_{W} + 2i G_{2}\right|
\right) \;.
\label{gnd}
\end{equation}
After a time span $\tau_{\perp}$ the coupling to the current has thus
reduced the spin state to a mixture, the charge specific density
matrix being diagonal in the measurement basis. We note that there is
no simple relation between $\tau_{\perp}$ and the decoherence time
$\tau_{d}$ in Eq.~(\ref{bvd}). Generally, $\alpha_{n}$ is more robust
relative to decoherence than $\sum_{n}\alpha_{n}$.

Although the characteristic times in Eqs.~(\ref{frd}), and (\ref{gnd})
depend on different combinations of parameters of the model, their
ratio seems to be rather universal and we expect quite generally that
\begin{equation}
\frac{\tau_{\perp}}{\tau_{z}}
=
c\;\tanh^2\frac{V}{2T}
\label{mnd}
\end{equation}
where $c$ is a constant of order unity.\cite{ratio}

We have identified the characteristic time scales in the problem, and
found their dependences on voltage and temperature.  The three time
scales describe the coherent dynamics of the spin under different
experimental conditions.  When the charge state of the junction is
left unobserved, the dynamics of the spin is given by the
unconditional spin density matrix, Eq.~(\ref{avd}), which decoheres on
the time scale, $\tau_d$, given in Eq.~(\ref{bvd}).  When the spin
evolution is conditioned on the charge state, the dynamics of the spin
is given by the charge specific density matrix.  In the shot noise
regime, $V\gg T$, spin-charge separation and the decoherence of the
charge specific density matrix happens on the same time scale,
$\tau_{\perp}\sim \tau_{z}$, and the tunnel junction provides a
projective measurement of the spin after a time span of this order.
For small voltages, $V\ll T$, when $\tau_{\perp}\ll \tau_{z}$, the
tunnel junction provides a projective measurement of the spin only
after the long time span $ \tau_{z}$.

To study the time evolution of the charge specific spin density matrix
in detail we turn to analyze the obtained analytical results.

\section{``Watching'' a quantum measurement}
\label{watch}

We shall now consider the time evolution of a measurement of the state
of the spin performed by the tunnel junction.  The charge specific
density matrix is presented in the Pauli basis
\begin{equation}
\hat{\rho}_{n}(t)=\frac{p_{n}(t)}{2}\left( \hat{1}+\mathbf{s}_{n}(t)
\bm{\cdot} \hat{\bm{\sigma}}\right),
\label{eq:1}
\end{equation}
where $p_{n}(t)$ is the probability that $n$ electrons have passed
through the junction at time $t$ and the polarization or Bloch vector
$\mathbf{s}_{n}(t)$ indicates a point within the unit Bloch sphere.

The coupling between the spin and the tunnel junction is chosen to
have the spin structure given in Eq.~(\ref{w4cx}), and the parameters
$ v_{\bm{lr}}$ and $ w_{\bm{lr}}$ are in the following taken to be
real constants.  In that case $G_2$ vanishes; the ratio of the
conductivities is taken to be $G_{W}=0.01 G_{V}$.

Initially the junction is in the definite charge state,
$p_{n}(0)=\delta_{n,0}$, as depicted in figure \ref{F1} (d).  For
definiteness, we choose the spin initially in the pure state
corresponding the Bloch vector $\mathbf{s}_{0}(0)=(-1,0,0)$.  The
further evolution of Bloch vectors takes place only in the xz-plane,
and the charge specific spin density matrix can therefore be
represented by a line rising vertically from the point indicating the
location of the Bloch vector $\mathbf{s}_{n}(t)$. Its height (ending
in a black dot for visual clarity) represents the probability for the
transfer of $n$ charges, $p_{n}(t)$.  The initial state is thus
represented as depicted in figure \ref{F1} $(a)$.  At a later time
$t$, many vertical lines are present, representing the probabilities
for various charge transfers and their corresponding spin states as
depicted in figure \ref{F1} (b).  In fact, many Bloch vectors
$\mathbf{s}_{n}(t)$ will appear close to each other, and placing a
square grid over the xz-plane cross section several Bloch vectors,
differing from each other in charge number by units of one, can lie in
the same square.  For visual clarity, however, only one vertical line
in the figures will be associated with a square, and its height is the
sum of the individual probabilities $p_{n}$ belonging to the Bloch
vectors in the square.

We now turn to study the time evolution of the charge specific spin
density matrix, and start by considering the shot noise regime where
the junction is biased by a high voltage, $V\gg T$.

\subsection{Schottky limit}

\begin{figure*}
\subfigure[\(t=0\) ]{\includegraphics[ 
  width=0.3\textwidth,
  keepaspectratio]{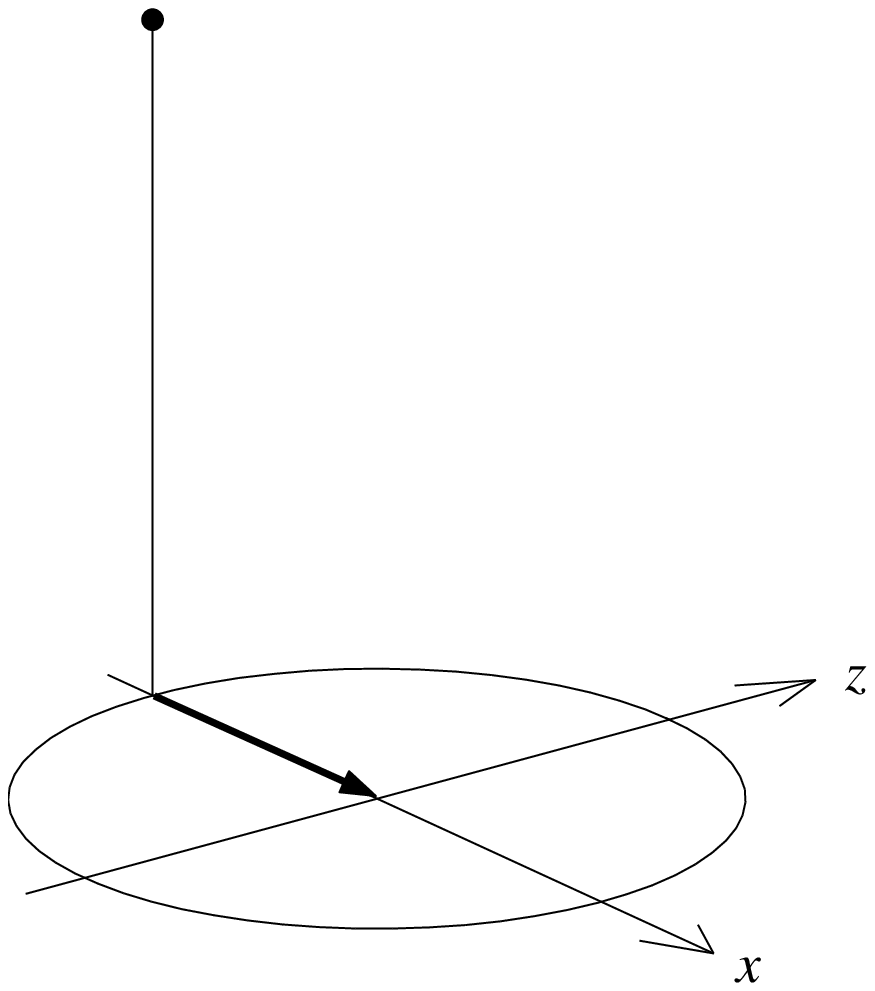}}
\subfigure[\(t= \tau_z/2\)]{\includegraphics[ 
  width=0.3\textwidth,
  keepaspectratio]{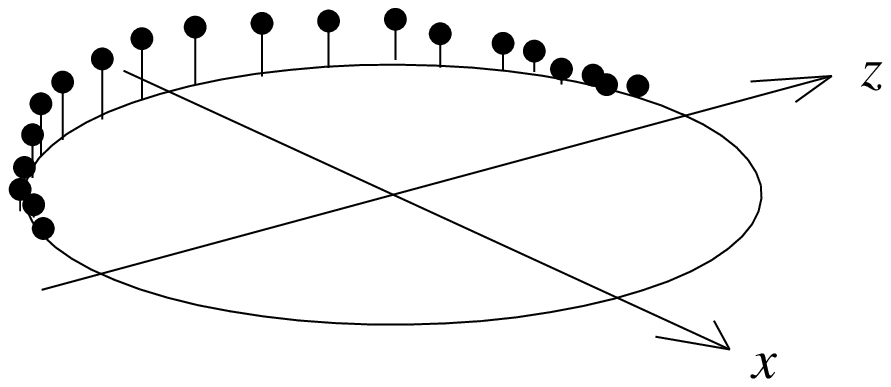}}
\subfigure[\(t=5 \tau_z\)]{\includegraphics[ 
  width=0.3\textwidth,
  keepaspectratio]{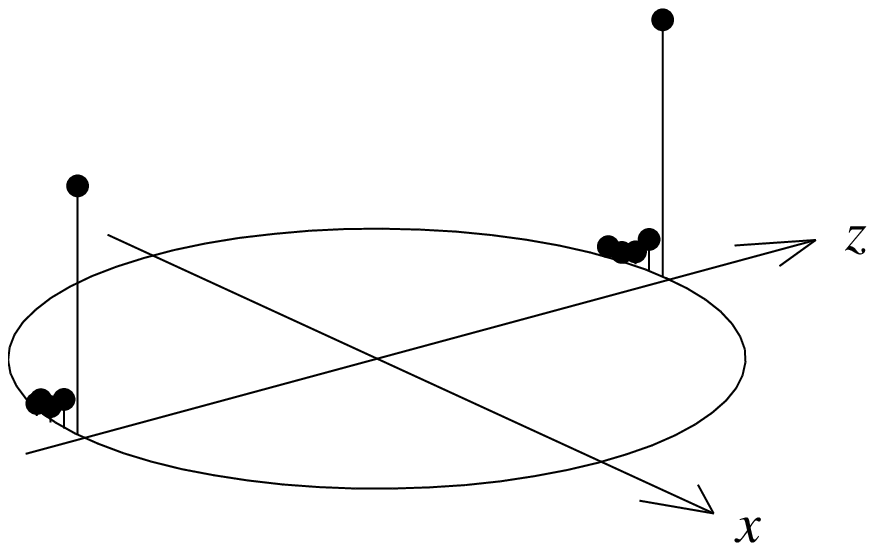}}

\subfigure[\(t=0\) ]{\includegraphics[ 
  width=0.3\textwidth,
  keepaspectratio]{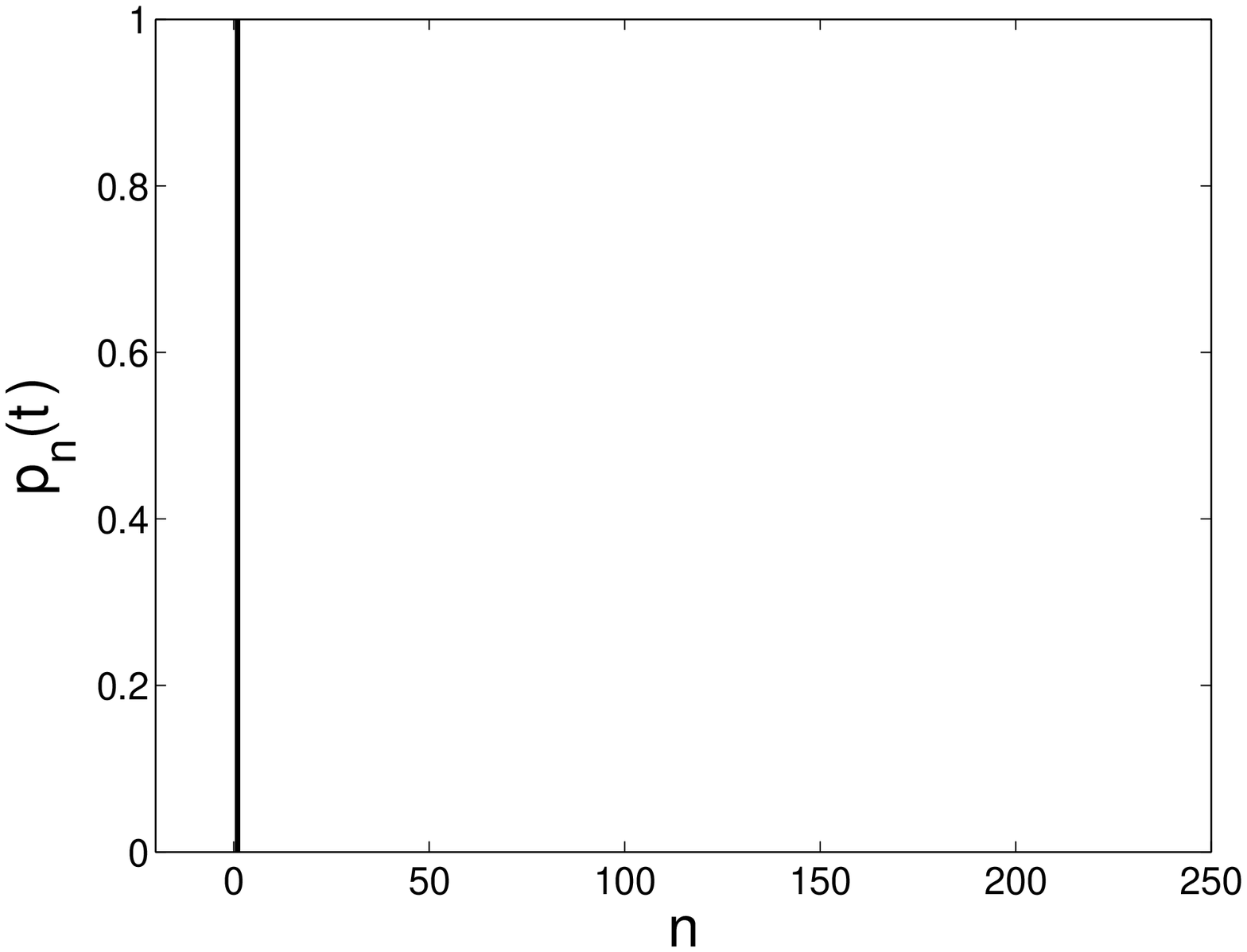}}
\subfigure[\(t=\tau_z/2\) ]{\includegraphics[ 
  width=0.3\textwidth,
  keepaspectratio]{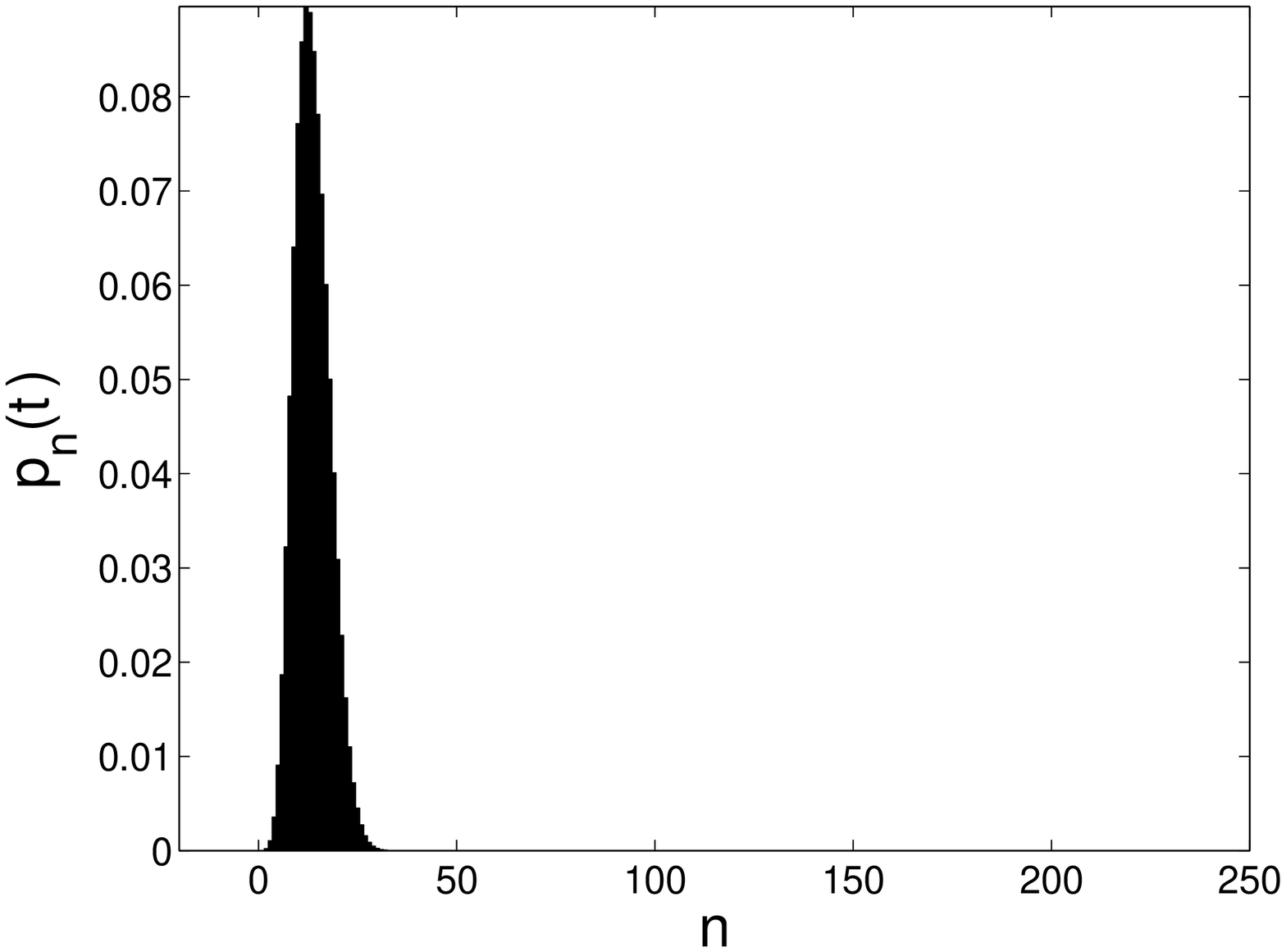}}
\subfigure[\(t= 5 \tau_z/\) ]{\includegraphics[ 
  width=0.3\textwidth,
  keepaspectratio]{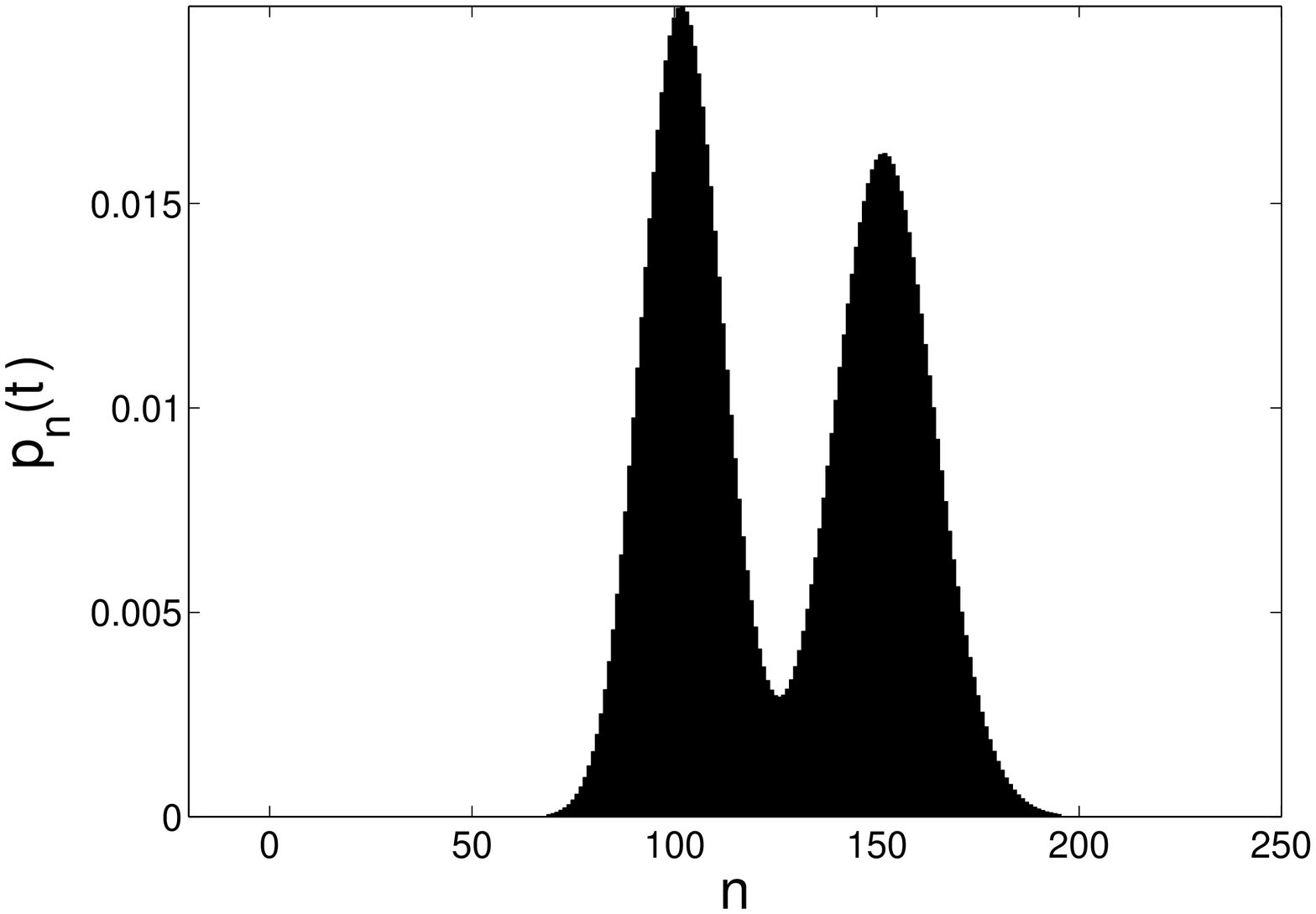}}
\caption{\label{F1} Spin-charge evolution in the high voltage case, $V
= 20 T$.  (a) Spin-charge representation of the initial state. The
arrow indicates the evolution of the unconditional spin density matrix
described by the Bloch vector shrinking along the $x$-axis decohering
to the mixture state $\mathbf{s}=(0,0,0)$ on time scale $\tau_{d} $.
(b) As time passes the Bloch spin density spreads along the unit
circle.  (c) At time $t=5\tau_{z}$ the Bloch spin density is located
at the pure states $\left|\uparrow\right\rangle $ and
$\left|\downarrow\right\rangle $.  Simultaneously, the initial charge
distribution, depicted in figure (d), will due to tunnelling events
start to move in charge space with a peak determined by the average
number of charges transferred.  (e) At time $t = \tau_{z}/2$, the
charge distribution, $p_n= u_n + d_n$, is built equally of the charge
conditioned probabilities for the spin to be in state up or down.  (f)
At time $t=5\tau_{z}$, the charge distribution is spin separated into
a peak built solely by the charge conditioned probabilities for the
spin to be in state up and a charge probability peak build by the
probabilities for the spin to be in state down. A projective spin
measurement has been performed by the tunnel junction after an average
of roughly hundred electrons have tunnelled.}  
\end{figure*}

The evolution of the Bloch spin density for the case where the voltage
is larger than the temperature, $V = 20 T$, is presented in figure
\ref{F1} (b) and (c).  At time $ t=\tau_{z}/2 \simeq \tau_{\perp}/2$,
the Bloch spin density is spread along the unit circle as depicted in
figure \ref{F1} (b) (the corresponding charge numbers increasing in
the clockwise direction).  The visualization shows that during the
evolution, the charge specific density matrix stays pure.  This is
characteristic of the shot noise regime, where the charge transfer
statistics is Poissonian, though not universal since to some extent
dependent on the choice of constant tunnelling matrix elements.  At
this time, $ t=\tau_{z}/2$, where on the average roughly ten electrons
have tunnelled, the charge probability distribution, $p_n= u_n + d_n$,
depicted in figure \ref{F1} (e), is built equally of the charge
conditioned probabilities for the spin to be in state up or down.  At
a larger time, $ t =5\tau_{z}$, the Bloch spin density distribution,
depicted in figure \ref{F1} (c), has split and is located near either
of the Bloch vectors \textbf{$\mathbf{s}_{\uparrow}=(0,0,1)$} or
\textbf{$\mathbf{s}_{\downarrow}=(0,0,-1)$}, corresponding to the spin
up and down states, respectively.  At this time, where on the average
roughly one hundred electrons have tunnelled, the charge probability
distribution, $p_n= u_n + d_n$, is spin separated into a charge
probability peak built solely by the probabilities for the spin to be
in state up and a charge probability peak built by the probabilities
for the spin to be in state down, and has the shape depicted in figure
\ref{F1} (f). The total probability in the two peaks both equal one
half, the probabilities initially for the spin to be in state up or
down.  The probability distributions for charge and spin have thus
through interaction come in one-to-one correspondence, and measuring
the charge state of the junction at times larger than $\tau_z$ is thus
a measurement of the state of the spin prior to interaction with the
tunnel junction.  The relative frequency with which a charge state in
either of the two peaks is realized is equal to the probability for
the corresponding spin state at the start of the measurement.  The
junction thus functions as a projective measuring device of the spin.
To be effective, we see from figure \ref{F1} (f), that on the average
hundred electrons must have tunnelled.

\subsection{Nyquist limit}

\begin{figure*}
\begin{center}\subfigure[\( t=\tau_\bot\)]{\includegraphics[ 
  width=0.3\textwidth,
  height=5cm,
  keepaspectratio]{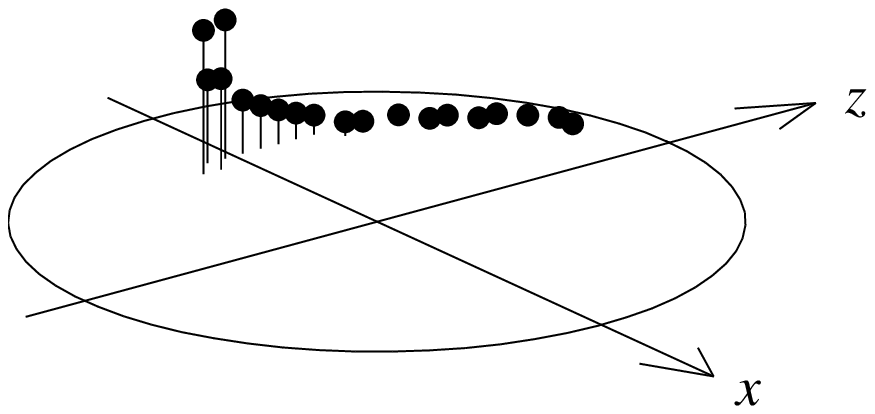}}\subfigure[\(
  t=\tau_z/2 \simeq 50 \tau_\bot\)]{\includegraphics[ 
  width=0.3\textwidth,
  height=5cm,
  keepaspectratio]{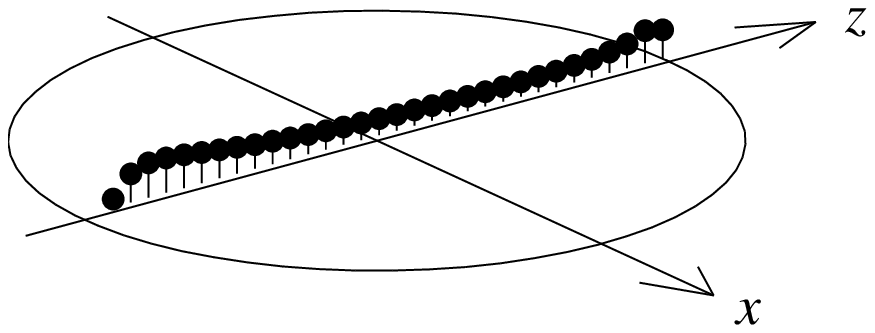}}\subfigure[\(t=5
\tau_z\)]{\includegraphics[ 
  width=0.3\textwidth,
  height=5cm,
  keepaspectratio]{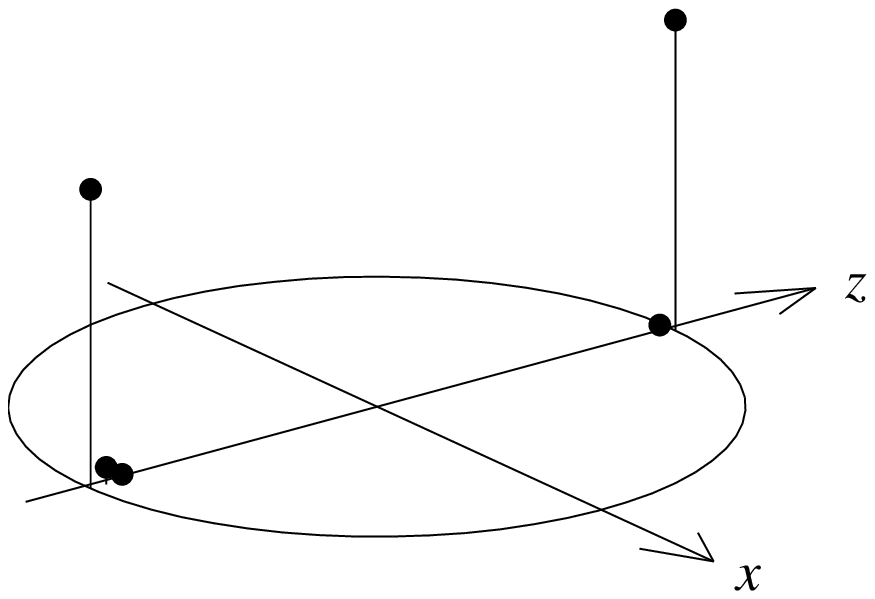}}\end{center}

\subfigure[\(t=\tau_\bot\) ]{\includegraphics[ 
  width=0.3\textwidth,
  keepaspectratio]{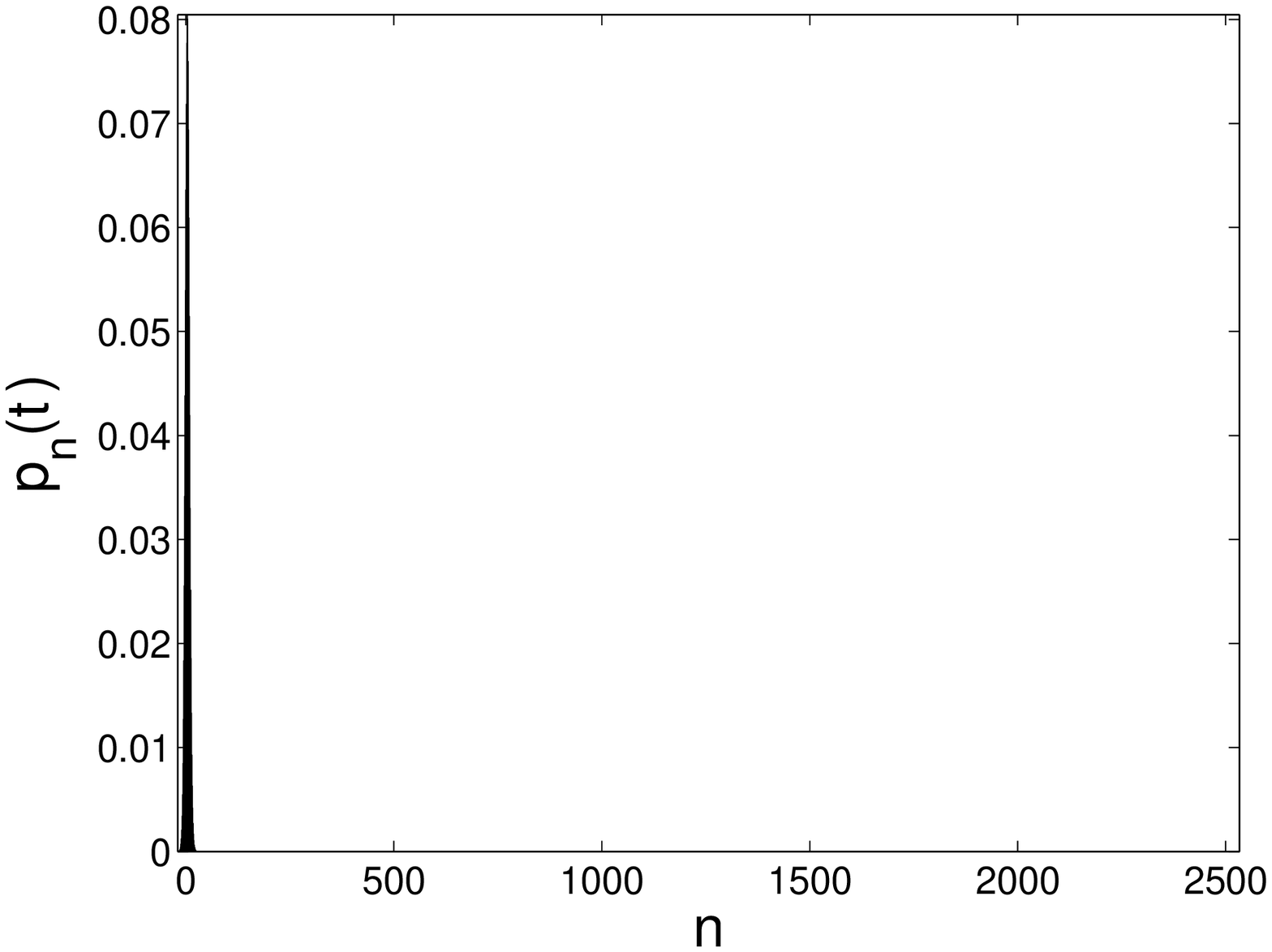}}\subfigure[\(t=\tau_z/2
  \simeq 50\tau_\bot\)
 ]{\includegraphics[ 
  width=0.29\textwidth,
  keepaspectratio]{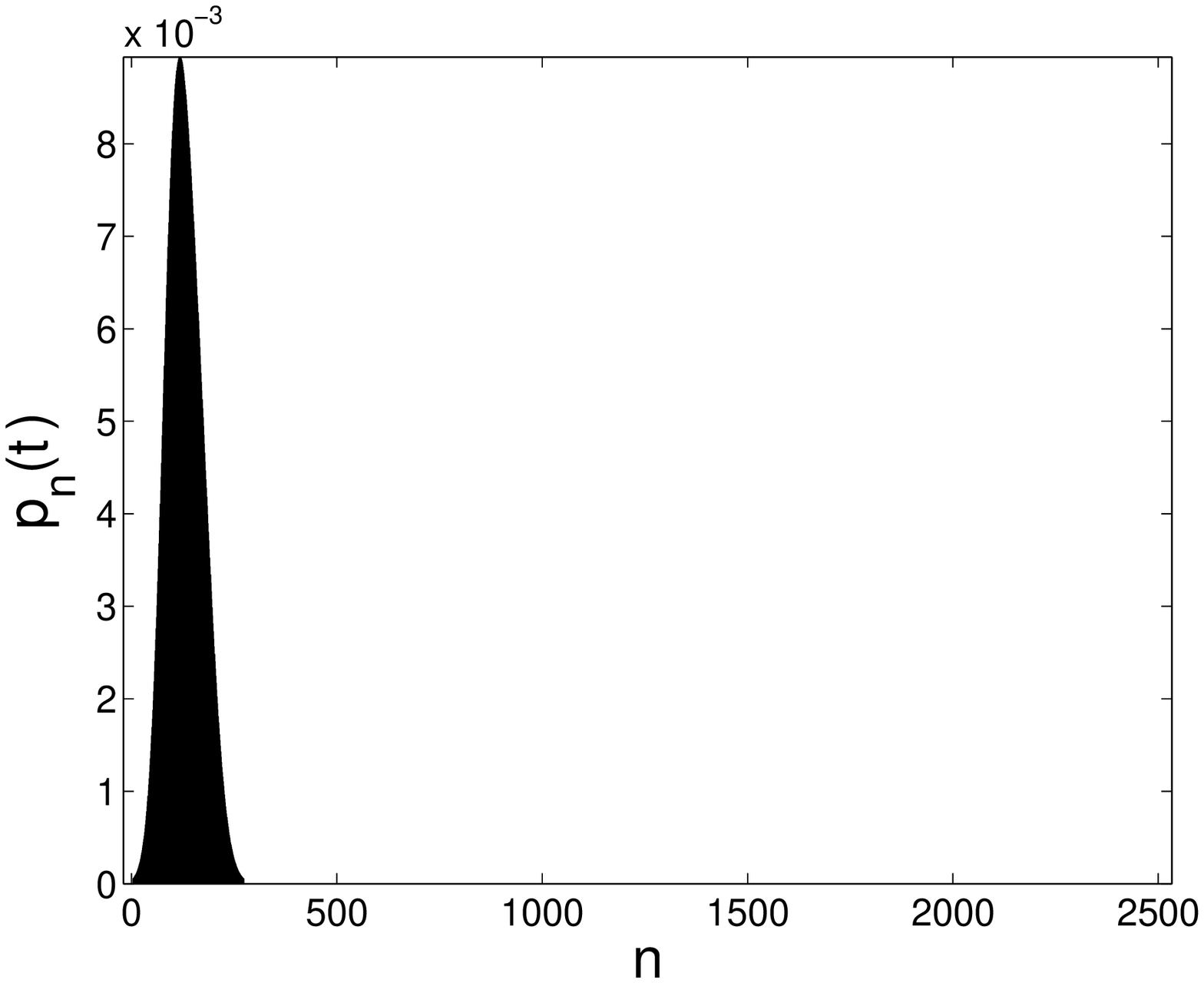}}\subfigure[\(t=  5 \tau_z
\) ]{\includegraphics[ 
  width=0.3\textwidth,
  keepaspectratio]{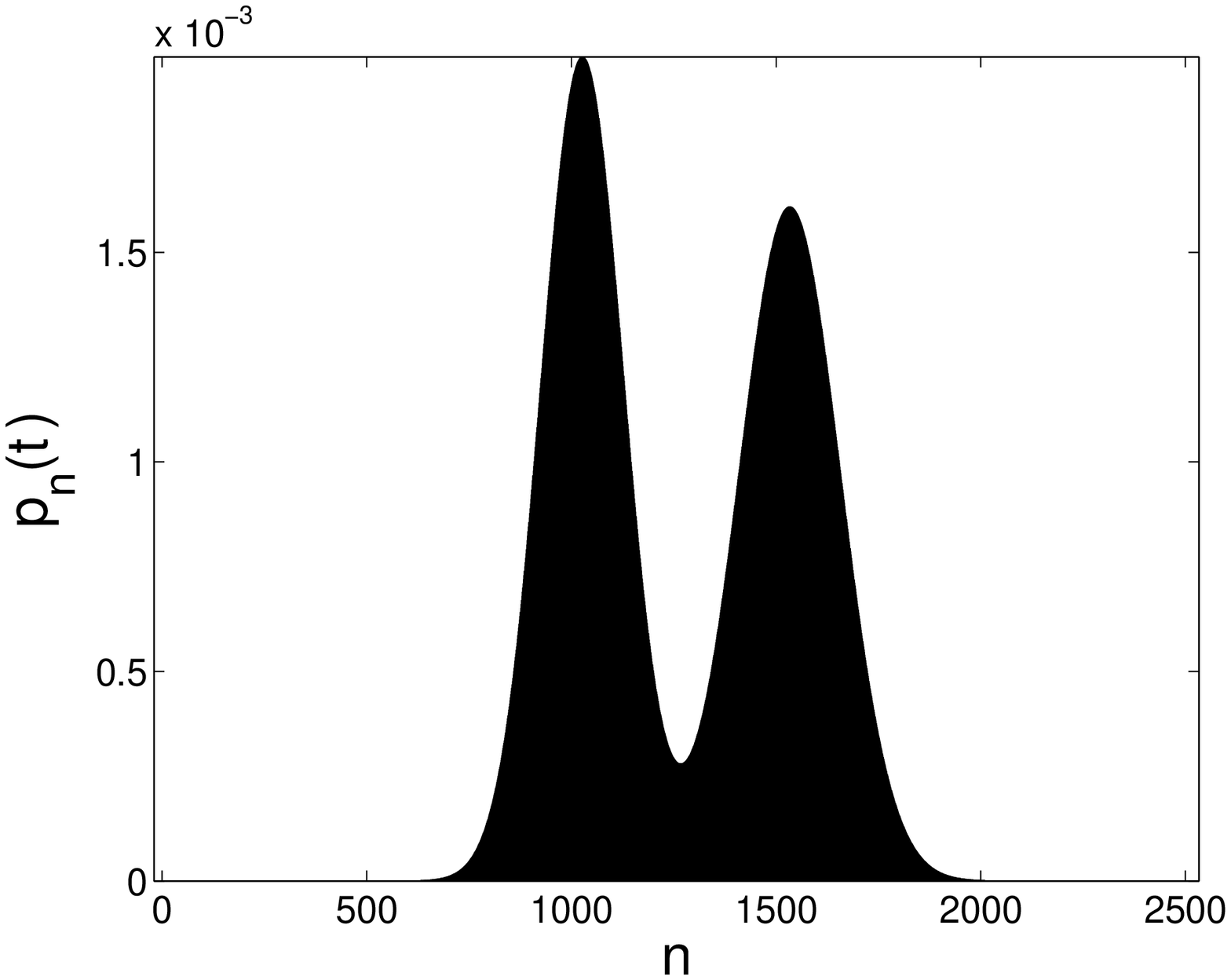}}

\caption{\label{F2}(a) Spin-charge evolution in the low voltage case,
$V = 0.2 T$.  (a) At time $t= \tau_\bot$, the x-component of the Bloch
vectors have started to decay, and the charge specific spin density
matrix is no longer pure.  (b) At the much larger time $t=\tau_{z}/2$,
$\tau_{z} \simeq 100 \tau_{\perp}$, the Bloch vectors are concentrated
on the z-axis, and the charge specific spin density matrix has evolved
into a mixture.  (c) At time $t=5\tau_{z}$, the Bloch spin density is
located at the pure states $\left|\uparrow\right\rangle $ and
$\left|\downarrow\right\rangle $.  (d) At time $t= \tau_\bot$, only an
average of hundred charges has tunnelled, and the charge distribution,
$p_n= u_n + d_n$, is built equally of the charge conditioned
probabilities for the spin to be in state up or down.  (e) At time
$t=\tau_z/2 \simeq 50 \tau_\bot$, the charge specific spin up and down
distributions, $u_n$ and $d_n$, have started to charge separate, but
not enough to significantly distort the total charge distribution.
(f) At time $t=5\tau_{z}$, the charge distribution is spin separated
into a peak built solely by the charge conditioned probabilities for
the spin to be in state up and a charge probability peak built by the
probabilities for the spin to be in state down. A projective spin
measurement has been performed by the tunnel junction after an average
of roughly one thousand electrons have tunnelled.  } 
\end{figure*}

Next we study the evolution of the charge specific spin density matrix
for the case where the voltage is smaller than the temperature, $V =
0.2 T$. The initial spin and charge states are chosen as in the high
voltage case depicted in figure \ref{F1} (a) and (d), so initially the
Bloch spin density is concentrated near the initial Bloch vector,
$\mathbf{s}=(-1,0,0)$.  However at larger times, the evolution of the
Bloch spin density is quite different from the high voltage case
proceeding in two steps as noticed from figure \ref{F2} (a).  At time
$\tau_{\perp}$, the $x$-component of the Bloch vectors,
$\mathbf{s}_{n}(\tau_{\bot})$, have started to decay, and the charge
specific spin density matrix is no longer pure. At this time on the
average only roughly five electrons have tunnelled according to figure
\ref{F2} (d).  At the much larger time $t=\tau_{z}/2$, $\tau_{z}
\simeq 100 \tau_{\perp}$, it is seen from figure \ref{F2} (b), that
the Bloch vectors are concentrated on the $z$-axis, the charge
specific spin density matrix has evolved into a mixture.  At this
time, where on the average roughly one hundred electrons have
tunnelled, the charge specific spin up and down distributions, $u_n$
and $d_n$, have started to charge separate, but not enough to
significantly distort the total charge distribution as evident from
figure \ref{F2} (e).  At the later time $t = 5\tau_{z}$, the Bloch
spin density is again separated and located at the sites corresponding
to the spin up and down states as depicted in figure \ref{F2} (c), and
the charge probability distribution is again spin separated, and the
total probability in the two peaks equals one half, the probabilities
initially for the spin to be in state up or down.  A projective spin
measurement has again been performed by the tunnel junction.  To be
effective, we see from figure \ref{F2} (f), that on the average
roughly one thousand electrons must have tunnelled.

The evolution of the initial state is seen to take place in two steps,
governed by the time scales $\tau_{\bot}$ and $\tau_{z}$. First the
pure initial state, $\mathbf{s}=(-1,0,0)$, decays into a mixture where
the Bloch vectors have no x-component, $\mathbf{s}=(0,0,s_{n}^{z})$,
(no off-diagonal elements in the charge specific density matrix).  The
mixture is then purified in the further time evolution, before the density 
matrix finally approaches the pure states $\left|\uparrow\right\rangle $ and 
$\left|\downarrow\right\rangle $ on the time scale $\tau_{z}$.

\section{Measurement time}\label{measure}

In the measuring scheme under consideration, information about the
spin state is obtained from a charge measurement. In the long time
limit, the charge probability distribution acquires two distinct peaks
(as seen in Figures \ref{F1} (f) and \ref{F2} (f)), and the outcome of
a charge measurement is almost always in the vicinity of
$Q_{\uparrow}$ {\em or} $Q_{\downarrow }$, $Q_{\uparrow, \downarrow }
= en_{\uparrow, \downarrow }$ being the positions of the
peaks. Ideally, one reads off the $z$-projection of the spin,
$\uparrow$ or $\downarrow $, from the charge outcome $Q_{\uparrow}$ or
$Q_{\downarrow }$. At any finite time, this scheme may produce
errors. The intrinsic error mechanism is electrical noise due to which
the peaks have a finite width and, therefore, partially
overlap. Additionally, the spin state is not pure at finite times,
spin purification occurring only asymptotically.  Also, errors may
occur due to a finite resolution $\Delta Q_{D}$ of the charge
detector.  The time span, $t_{m}$, needed to make a reliable
measurement, is defined by the condition that the information gained
from the charge measurement suffices to identify the $z$-projection of
the spin after the measurement.  First, we shall consider the case of
an ideal charge detector, before taking into account the finite
resolution.  \paragraph{An ideal charge detector,}generates certain
discrete values, $n$, as the outcome of the charge measurement with
the probability $p_{n}= {\rm{Tr}}\;\hat{\rho}_{n}$, and leaves the
spin in the state described by the density matrix proportional to
$\hat{\rho}_{n}$. With $\hat{\rho}_{n}$ of the form in
Eq.~(\ref{skd}), the probability $p_{n}$ equals $p_{n}= u_{n}+ d_{n}$,
and
\begin{equation}
p_{n}^{\uparrow}=\frac{u_{n}}{u_{n}+d_{n}}\quad,\quad 
p_{n}^{\downarrow}=\frac{d_{n}}{u_{n}+d_{n}}
\label{hvd}
\end{equation} 
are the probabilities for the spin to be in the up or down states for
a given {\em observed} $n$.

Asymptotically, the distributions $u_{n}(t)$ and $d_{n}(t)$ do not
overlap in accordance with Eq.~(\ref{wnd}). Therefore, only one of the
diagonal elements, $u_{n}$ or $d_{n}$, remains finite and the spin is
left in either the up or down state.  At long times, the measurement
is perfectly projective.

To quantify the efficiency of the measurement at finite times, we take
an approach well known in the context of information theory (see
e.g. ref.  \onlinecite{NieChu00}).  Considering the outcomes of the
joint measurement of the charge $n$ and spin $\sigma = \uparrow,
\downarrow $, as random variables, the probability for the outcome
$(n, \uparrow)$ (or $(n, \downarrow )$) is given by $u_{n}$ (or
$d_{n}$).  The joint entropy, $H(\sigma,n)$, equals
\begin{equation}
 H(\sigma,n)=
-\sum_{n}\left(u_{n}\log_{2}u_{n}+d_{n}\log_{2}d_{n}\right),
\label{ivd}
\end{equation}
and the entropy $H(n)$ of the charge variable $n$ is,
\begin{equation}
 H(n)=-\sum_{n}p_{n}\log_{2}p_{n} \;.
\label{jvd}
\end{equation}

\begin{figure}
\begin{center}\includegraphics[ 
  width=.75\columnwidth,
  keepaspectratio]{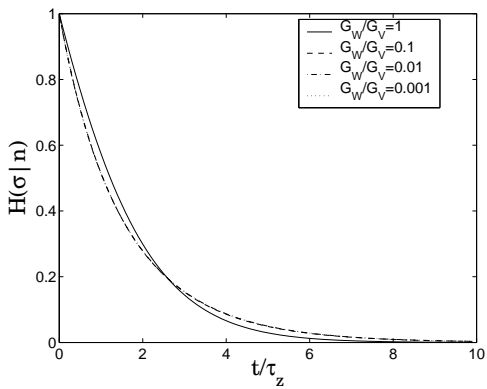}\end{center}
\caption{\label{Shannon-v10}The conditional entropy for the high
voltage case, $V=20T$, as a function of time for the initial state
$\left|\psi\right\rangle =1/\sqrt{2}\left(\left|\uparrow\right\rangle
+\left|\downarrow\right\rangle \right)$,
and four different choices of dimensionless conductance.  The
conditional entropy is seen to decrease monotonically with a time
constant of the order of $\tau_{z}$. The entropy curves for
$g_{W}=0.1,\,0.01\,\text{and\,0.001}$ all collapse onto one curve.}
\end{figure}

The conditional entropy,
$H(\sigma |n)= H(\sigma ,n)- H(n)$, is a measure of the information
gained about the spin from a charge measurement. Combining
Eq.~(\ref{ivd}) and Eq.~(\ref{jvd}), we obtain
\begin{equation}
 H(\sigma|n)=-\sum\limits _{n}p_{n}\left(p_{n}^{\uparrow}
\log_{2}p_{n}^{\uparrow}+p_{n}^{\downarrow}\log_{2}p_{n}^{\downarrow}\right)\;.
\label{kvd}
\end{equation}
This quantity changes from the initial value,
$-\left(u_{0}\log_{2}u_{0}+d_{0}\log_{2}d_{0}\right)$, to zero at
large times where $p_{n}^{\uparrow,\downarrow}$ is either zero or 1.
At intermediate times it is a measure of the error in the predicted
spin state after the charge measurement.  The time dependence of the
conditional entropy for the initial state of the spin in the
$x$-direction is shown in Fig. \ref{Shannon-v10}.  The curve shows
exponential-like decrease of the conditional entropy from the initial
value, one bit, to the value zero.

\begin{figure}
\begin{center}\includegraphics[ 
  width=.75\columnwidth,
  keepaspectratio]{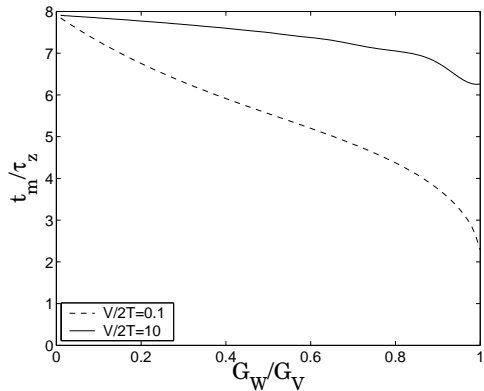}\end{center}
\caption{\label{measure-time}
The measurement time $t_{m}$, defined as the time when the entropy has
dropped to one percent, as a function of the coupling strength
$G_{W}/G_{V}$, for the initial state $\left|\psi\right\rangle
=1/\sqrt{2} \left(\left|\uparrow\right\rangle
+\left|\downarrow\right\rangle \right)$, and high and low
bias voltage.  It is seen that for $\tau_{z}$ varying by two orders of
magnitude, $t_{m}$ roughly scales with $\tau_{z}$ as given by
Eq. (\ref{frd}). } \end{figure}

For the case of an ideal charge detector, we define the measurement
time $t_{m}$ as the time when the spin entropy decreases below a
chosen threshold determined by the required fidelity of the
measurement.  In Fig. \ref{measure-time}, we show the measurement time
for various parameters with the entropy threshold (arbitrarily) chosen
as 0.01.  The main conclusion is that the intrinsic measurement time
is of order $\tau_{z}$ as given in Eq.~(\ref{frd}), $t_{m}\sim
\tau_{z} $.

The most favorable condition for the accuracy of the measurement is
the shot noise regime where the charge noise is relatively small.  As
a rough estimate of the measuring time, $t_{m}\sim \tau_{z} $, in the
shot noise regime we obtain from Eq.~(\ref{frd})
\begin{equation}
t_{m}^{-1} \sim \gamma  \frac{eU}{\hbar }
\quad,\quad \gamma = 
\frac{8G_{1}^2}{G_{V}+G_{W}}
\label{mvd}
\end{equation}
where $U$ is the voltage applied to the junction and 
$\gamma $
is a dimensionless constant.  For the tunnelling Hamiltonian approach
and the Markovian approximation to be justified, the constant $\gamma
$ should be small.  If the conducting modes are fully spin effective,
i.e., $v_{\bm{lr}}\sim w_{\bm{lr}}$ in Eq.~(\ref{w4cx}), the constant
$\gamma$ is of the order of the conductivity of the junction in units
of $e^2/ \hbar $.  Although smallness of $\gamma $ is needed for the
derivation, the theory is expected to be qualitatively applicable even
for $\gamma \sim 1$. From this one concludes that the intrinsic
quantum bound on the measurement time is given by the inequality, $
t_{m} \gtrsim \hbar / eU $.

The separation, $\Delta Q = |Q_{\uparrow}- Q_{\downarrow }|$, of the
two spin orientations in charge space by the time of measurement, can
be evaluated as $\Delta Q \sim 2|G_{1}U|t_{m}$.  Using the estimate in
Eq.~(\ref{mvd}),
\begin{equation}
\frac{\Delta Q}{e}  \sim \frac{G_{V}+ G_{W}}{4|G_{1}|}
\label{pvd}
\end{equation}
for any voltage (in the shot noise regime). 

\paragraph{A real charge detector,} may have a resolution lower than
$\Delta Q$ given by Eq.~(\ref{pvd}). In this case, the time of
measurement is estimated from the condition that the distance in the
charge space between the peaks corresponding to opposite spin
orientations, exceeds the detector resolution, i.e., $\Delta Q(t_{m})
{}> \Delta Q_{D}$.  From this, we get for the lower bound on the
measurement time
\begin{equation}
t_{m}\gtrsim \frac{\Delta Q_{D}}{e} \frac{e^2}{2 \hbar |G_{1}|}
\frac{\hbar }{eU}
\; .
\label{qvd}
\end{equation}
In this case the time of measurement essentially depends on the
resolution of the charge measuring device.

In recent experiments involving quantum point contacts coupled to
two-level systems, realized by coupled quantum dots, the typical
voltage bias was in the $mV$-range and the Rabi frequency of the order
of $10^{10}Hz$.\cite{HayFujChe03,DicLynJoh03} In accordance with the
estimate of the measurement time, the scheme studied in the present
paper can be realized in currently studied nanostructures.

\section{Summary and conclusion}\label{sum}

We have derived the equation of motion, Eq.~(\ref{hjd}), for the
charge specific density matrix for a quantum system interacting with
an environment, the role of which is played by a tunnel junction
connecting two electron reservoirs. In our approach we use the density
matrix technique with the environment degrees of freedom partially
traced out: We keep track of the charge distribution between the
reservoirs and derived the master equation for the charge specific
density matrix of the system, $\hat{\rho}_{n}$, $n$ being the charge
variable. The interaction with the environment is through the
dependence of the tunnelling amplitudes on the state of the system.
This dependence being specified, the derived master equation,
Eq.~(\ref{hjd}), is general and can be applied to any system.  The
charge distribution is a collective variable of the environment, and
we have shown that to lowest order in the tunneling it can be treated
classically.

In our analysis of the measurement of a spin (two-level system), the
charge variable plays the role of the pointer coordinate in von
Neumann's general theory of quantum measurement.\cite{Neu32} At the
conceptual level, the measurement scheme works as follows: One first
disconnects the junction, preparing thereby the environment in a
certain initial charge state, say $n=0$. To measure the spin wave
function $|\psi_{0}\rangle = \alpha |\uparrow \rangle + \beta
|\downarrow \rangle $ at time $t=0$, the spin sensitive tunnelling,
Eq.~(\ref{w4cx}), is switched on at $t=0$ together with the voltage
$U$ and kept till the instant $t=t_{m}$, the measurement time
discussed in Section \ref{measure}.  Then, the junction is again
disconnected, and the transferred charge $Q=en$, the change in the
charge of one of the electrodes, is examined. The transferred charge
$Q$ is random due to noise, both classical and quantum. However, its
probability distribution at the instant $t=t_{m}$ is concentrated in
two well-separated peaks located at $Q_{\uparrow}$ and $Q_{\downarrow
}$ (as seen in Figures \ref{F1} (f) and \ref{F2} (f)).  The frequency
of occurrence of transferred charge $Q$ in the vicinity of the
corresponding peaks, gives the probabilities $|\alpha |^2$ and $|\beta
|^2$. The measurement is projective: An observation of the charge
state in a state around $Q_{\uparrow}$ (or $Q_{\downarrow }$) ensures
that the spin after the measurement is in the pure state $|\uparrow
\rangle $ (or $|\downarrow \rangle $).

In the shot noise regime, which is the most favorable for performing
the measurement, the measurement time and the voltage $U$ are related
as $eU\, t_{m} \sim \hbar / \gamma $ where $\gamma$ (see
Eq.~(\ref{mvd})) is the effective coupling constant. For our theory to
be valid, the coupling constant must be small. However, one may expect
that the estimate holds even for $\gamma \sim 1$, giving the ultimate
quantum bound, $eU\, t_{m} \gtrsim \hbar$, for the time duration of
the interaction needed in order for the tunnel junction to function as
a measuring device. On the other hand, the duration must be short on
the scale of the inverse Rabi frequency $\Omega $ for the measurement
to be ``instantaneous''. For the typical Rabi frequencies $\Omega \sim
10^{10}Hz$,\cite{HayFujChe03,DicLynJoh03} the measurement becomes
``instantaneous'', that is $\Omega t_{m}\ll 1$, provided the amplitude
of the voltage pulse $U$ is above the $ mV$ level.

The application of the charge projection method to a tunnel junction,
a semi-realistic model of a nanodevice, allows us to study questions
regarding quantum measurement, and we have considered the emergence of
a projective spin measurement.  We have obtained explicit expressions
for the characteristic times for decoherence and spin state
purification, and visualized the emergence of a projective spin
measurement.  The model allowed a detailed study of the purification
of the charge specific spin density matrix.  The spin measurement by
the tunnel junction is quite analogous to the spin measurement in the
Stern-Gerlach experiment, except that the interaction mechanism of
course is rather different, being due to interaction with reservoirs
of charged fermions.  This gives another example of a measuring scheme
showing that there is no need to postulate the classicallity of any
variable and no ``wave function collapse'' needs to be invoked. The
transmission of electrical noise from the tunnel junction to the
quantum object accomplishes the projective measurement.  Amplification
to the classical level and the emergence of a projective measurement,
have thus been illustrated using a model of a nanodevice.

The use of charge projectors to study charge kinetics originated in
the context of counting statistics.\cite{SheRam03} The derived master
equation for the charge specific density matrix shows that the method
of charge projectors leads to a useful application of counting
statistics.  Finally, we note that the obtained master equation allows
the study of the charge specific dynamics of an arbitrary quantum
object coupled to a tunnel junction.  The presented method is
therefore suitable for attacking problems in a variety of fields, such
as for example quantum electro-mechanical effects in
nanostructrures.\cite{KHoRamSheWab04}

\acknowledgments
We thank Dr.~Denis Khomitsky for helpful discussions.  Part of the
paper was completed during a visit of one of us (A. S.) to Argonne
National Laboratory and the hospitality extended during the visit is
greatly appreciated.  This work was supported by The Swedish Research
Council.  

\appendix

\section{Derivation of equation of motion}\label{derv}

In this appendix, the equation of motion for the charge specific
density matrix, $\hat{\rho }_{n}$, is derived for the case where
tunnelling events are rare, i.e., to lowest order in the tunnelling.
It is convenient to use the Heisenberg picture
\begin{equation}
\rho'
 = 
e^{i \hat{H}_{S}t}
\,\rho\,
e^{-i \hat{H}_{S}t} 
\label{gdd}
\end{equation} 
where the density matrix evolves only due to the dynamics of the 
electrons
\begin{equation}
i \dot{\rho }'
 = [H', \rho']
\label{ydd}
\end{equation}
and the Hamiltonian in the Heisenberg picture is
\begin{equation}
H'(t) = H_{l} + H_{r} + H_{T}'(t)
\label{kdd}
\end{equation}
where $H_{T}'(t) = {\cal T}(t) + {\cal T}^{\dagger}(t)$ and
\begin{eqnarray}
{\cal T}(t)
=
\sum\limits_{\bm{l}, \bm{r}} 
\hat{T}_{\bm{l}\bm{r}}(t) 
c_{\bm{l}}^{\dagger}c_{\bm{r}}
,\nonumber\\ 
\hat{T}_{\bm{l}\bm{r}}(t)
=
e^{i \hat{H}_{S}t}
\hat{T}_{\bm{l}\bm{r}}
e^{-i \hat{H}_{S}t} .
\label{idd}
\end{eqnarray}

The equation of motion in the Heisenberg picture for the charge
specific density matrix of the system coupled to the junction can be
written on the form
\begin{equation}
 \dot{\hat{\rho }}_{n}'(t)=
\sum\limits_{\bm{l}\bm{r}}
\left(
\hat{T}_{\bm{l}\bm{r}}(t)
\hat{A}_{\bm{l}\bm{r}}^{(n)}(t)
+ 
\hat{T}_{\bm{r}\bm{l}}(t)
\hat{B}_{\bm{r}\bm{l}}^{(n)}(t)
\right) + h.c.
\label{wdd}
\end{equation}
where the time dependent operators $\hat{A}$ and $\hat{B}$ are
\begin{eqnarray}
\hat{A}_{\bm{lr}}^{(n)}(t)= 
\frac{1}{i}
{\rm{Tr}}_{el}\;
\left(
 c_{\bm{l}}^{\dagger}c_{\bm{r}}\rho'(t){\cal P}_{n}
 \right)
,\nonumber \\ 
\hat{B}_{\bm{lr}}^{(n)}(t)= 
\frac{1}{i}
{\rm{Tr}}_{el}\;
\left(
 c_{\bm{r}}^{\dagger}c_{\bm{l}}\rho'(t){\cal P}_{n}
 \right) .
\label{xdd}
\end{eqnarray}
A hat indicates operators with respect to the degrees of freedom of
the system coupled to the junction.

To obtain, e.g. $\hat{A}$, one uses the equation of motion,
Eq.~(\ref{ydd}), to obtain
\begin{equation}
 \dot{\hat{A}}_{\bm{lr}}(t) - i \omega_{\bm{lr}} \hat{A}_{\bm{lr}}(t) =
-{\rm{Tr}}_{el}\;
\left(
 c_{\bm{l}}^{\dagger}c_{\bm{r}}[H_{T}'(t),\rho'(t)]{\cal P}_{n}
 \right)
\label{4dd}
\end{equation}
where the term containing $\omega_{\bm{lr}}\equiv
\varepsilon_{\bm{l}}- \varepsilon_{\bm{r}}$ originates from the
commutator with the electrode Hamiltonian, $H_{0}=H_{l}+ H_{r}$.  We
must therefore consider the inhomogeneous term on the right of
Eq.~(\ref{4dd}). According to the relations Eq.~(\ref{jdd}) and
Eq.~(\ref{ndd}) it can be rewritten on the form
\begin{eqnarray}
&&{\rm{Tr}}_{el}
\left(
 c_{\bm{l}}^{\dagger}c_{\bm{r}}[({\cal T}^{\dagger}+{\cal T})\rho']
{\cal P}_{n}
 \right)     
 =  {\rm{Tr}}_{el}\;
\left(
 c_{\bm{l}}^{\dagger}c_{\bm{r}}{\cal T}{\cal P}_{n-2}\rho'{\cal P}_{n}
 \right)
             \label{6dd} \nonumber\\   
      &&-  
{\rm{Tr}}_{el}\;
\left(
 c_{\bm{l}}^{\dagger}c_{\bm{r}}{\cal P}_{n-1}\rho'{\cal P}_{n+1}
{\cal T}^{\dagger}
 \right)
+
{\rm{Tr}}_{el}\;
\left(
 c_{\bm{l}}^{\dagger}c_{\bm{r}}{\cal T}^{\dagger}{\cal P}_{n}\rho'{\cal P}_{n}
 \right)
              \label{6dd3}\nonumber \\  
     &   &
-{\rm{Tr}}_{el}\;
\left(
 c_{\bm{l}}^{\dagger}c_{\bm{r}}{\cal P}_{n-1}\rho'{\cal P}_{n-1}{\cal T}
 \right)
\label{6dd2}
\end{eqnarray}
where the time argument has been suppressed.  The first two terms are
charge ``far'' off-diagonal components of the density matrix like
${\cal P}_{m}\rho {\cal P}_{m\pm 2}$, and these terms can be neglected
since they are of higher order in the tunnelling matrix element and an
expression involving only the two charge diagonal components of the
density matrix results.  These charge diagonal terms, which have the
form
\begin{equation}
{\rm{Tr}}_{el}
(
 c_{\bm{l}}^{\dagger}c_{\bm{r}}{\cal T}^{\dagger}
{\cal P}_{n}\rho'{\cal P}_{n}
) 
= \sum\limits_{\bm{l'r'}}
\hat{T}_{\bm{r'l'}}
{\rm{Tr}}_{el}
(
 c_{\bm{l}}^{\dagger}c_{\bm{r}}
 c_{\bm{r}'}^{\dagger}c_{\bm{l}'} 
{\cal P}_{n}\rho'{\cal P}_{n}
 ),
\label{8dd}
\end{equation}
can be  approximated by the expressions 
\begin{eqnarray}
&{\rm{Tr}}_{el}\;
(
 c_{\bm{l}}^{\dagger}c_{\bm{r}}
 c_{\bm{r}'}^{\dagger}c_{\bm{l}'} 
{\cal P}_{n}\rho'{\cal P}_{n}
 ) 
=
\delta_{\bm{ll'}}
\delta_{\bm{rr'}}
f_{\bm{l}}(1- f_{\bm{r}}) \hat{\rho}_{n}'&
,\nonumber \\ &
\hat{\rho}_{n}' = 
{\rm{Tr}}_{el}\;
\left(
{\cal P}_{n}\rho'{\cal P}_{n}
 \right) &
\label{9dd}
\end{eqnarray}
where the $f$'s are the single particle energy distribution functions
for the electrodes. The distribution functions are taken independent
of the number of charges on the left electrode $n$, since the
electrodes are inexhaustible particle reservoirs.  The diagonality in
the electrode quantum numbers are justified by the fact that when
tunnelling is rare, superpositions of different single particle states
do not occur.

Approximating similarly the second charge diagonal term,
Eq.~(\ref{6dd2}), makes Eq.~(\ref{4dd}) acquire the form
\begin{equation}
 \dot{\hat{A}}_{\bm{lr}} - i \omega_{\bm{lr}} \hat{A}_{\bm{lr}}
=
-f_{\bm{l}}(1- f_{\bm{r}})\hat{T}_{\bm{rl}}\; \hat{\rho}_{n}'
+
f_{\bm{r}}(1- f_{\bm{l}}) \hat{\rho}_{n-1}'\hat{T}_{\bm{rl}} .
\label{bed}
\end{equation}
In the following we shall take as initial condition that tunnelling
first starts at time $t=0$, i.e., the junction is disconnected at
earlier times. The solution of Eq. (\ref{bed}) thus becomes
\begin{eqnarray}
\hat{A}_{\bm{lr}}(t)&=&
-
f_{\bm{l}}(1- f_{\bm{r}})
 \ssss{\hat{T}_{\bm{rl}}(t)\hat{\rho}_{n}'(t)}\nonumber \\
&&+
 f_{\bm{r}}(1- f_{\bm{l}})  
 \ssss{\hat{\rho}_{n-1}'(t)
 \hat{T}_{\bm{rl}}(t)
}
\label{ged}
\end{eqnarray}
where the notation
\begin{equation}
\ssss {X(t)}
\equiv 
\int\limits_{0}^{t}dt' e^{i
 \omega_{\bm{lr}} (t-t') + i\int\limits_{t'}^{t} dt'' V(t'') }
X(t')
\label{omj}
\end{equation}
has been used, and the feature that the junction may be biased by a
time dependent voltage, $V(t)$, is included.

The evaluation of $B$ in Eq.~(\ref{xdd}) is analogous and we obtain
(of course in the same approximation and for the same initial
condition)
\begin{eqnarray}
\hat{B}_{\bm{rl}}(t)&=&
-
f_{\bm{r}}(1- f_{\bm{l}})
 \ssssd
{\hat{\rho}_{n}'(t)\hat{T}_{\bm{lr}}^{\dagger}(t)}\nonumber \\
&&+
 f_{\bm{l}}(1- f_{\bm{r}})
 \ssssd{
\hat{T}_{\bm{lr}}^{\dagger}(t)\hat{\rho}_{n+1}'(t)
}
\label{qed}
\end{eqnarray}
where the dagger indicates hermitian conjugation of the system operators. 

Collecting the results, we obtain a non-Markovian master equation for
the charge specific density matrix:
\begin{eqnarray}
 \hat{\dot{\rho }}_{n}'(t)&=& \Lambda' \sss{\hat{\rho}_{n}'(t)} + 
{\cal D}_{+}'\sss{\hat{\rho}_{n+1}'(t) - \hat{\rho}_{n}'(t)}
\nonumber \\&+&
{\cal D}_{-}'\sss{\hat{\rho}_{n-1}'(t) - \hat{\rho}_{n}'(t)}
\label{2ed}
\end{eqnarray}
where the kernels  are
\begin{eqnarray}
&\Lambda'& \sss{\hat{\rho }} =
\sum\limits_{\bm{l}\bm{r}}
f_{\bm{l}}(1- f_{\bm{r}})
\left(
 \ssss{
\hat{T}_{\bm{lr}}^{\dagger}\hat{\rho}}
\hat{T}_{\bm{l}\bm{r}}
-\hat{T}_{\bm{l}\bm{r}} 
\ssss{\hat{T}_{\bm{lr}}^{\dagger}\hat{\rho}}\; \right) +h.c. \nonumber \\
&+&
\sum\limits_{\bm{l}\bm{r}}
 f_{\bm{r}}(1- f_{\bm{l}})
\left(
\hat{T}_{\bm{l}\bm{r}}
\ssss{\hat{\rho}
 \hat{T}_{\bm{lr}}^{\dagger}
}
-
 \ssss
{\hat{\rho}\hat{T}_{\bm{lr}}^{\dagger}}\hat{T}_{\bm{l}\bm{r}}
 \right)
 + h.c.   \label{yxed2}
\end{eqnarray}
and
\begin{equation}
{\cal D}_{+}'\sss{\hat{\rho} }=
\sum\limits_{\bm{l}\bm{r}}
 f_{\bm{l}}(1- f_{\bm{r}})
 \ssss{
\hat{T}_{\bm{lr}}^{\dagger}\hat{\rho }
}
\hat{T}_{\bm{lr}}
 + h.c.
\label{xyed}
\end{equation}
and
\begin{equation}
{\cal D}_{-}'\sss{\hat{\rho} }=
\sum\limits_{\bm{l}\bm{r}}
 f_{\bm{r}}(1- f_{\bm{l}})
\hat{T}_{\bm{lr}} 
\ssss{
\hat{\rho }\hat{T}_{\bm{lr}}^{\dagger}
} + h.c.
\label{xzed}
\end{equation}

The charge specific spin density matrix in the Heisenberg picture
evolves with a rate proportional to the electron tunnelling rate.  The
temporal non-locality of the kernels in Eq. (\ref{2ed}), however, is
independent of tunnelling, and instead depends on the quantum time
scale determined by temperature and voltage. In the limit where
tunnelling can be neglected on this time scale, the master equation
for the spin dynamics becomes Markovian since $\rho '$ can be taken
outside the bracket operation $\ssss{...}$ in
Eq.~(\ref{yxed2}-\ref{xzed}). We therefore finally arrive at the
Markovian master equation, Eq. (\ref{hjd}), however there displayed in
the Schr\"{o}dinger picture.

\section{One-component master equation}\label{onecomp}

The equations Eqs.~(\ref{fqd}) and (\ref{tkd}-\ref{wkd}), have the
structure
\begin{equation}
\dot{x}_{n} =  D_{x}x_{n}'' +  J_{x} x_{n}' 
\quad,\quad x_{n}(t=0)= x_{0}\,\delta_{n,0} .
\label{xkd}
\end{equation}
The equation can be solved by Fourier transform,
\begin{equation}
x_{n}(t) = x_{0}
\int\limits_{0}^{2\pi}\frac{d\varphi}{2\pi} \, 
e^{( 2D_{x} (\cos \varphi -1) +i J_{x} \sin \varphi )t + i \varphi n}
  \label{3kd} 
\end{equation}
giving
\begin{equation}
x_{n}(t) = x_{0}
\left(\sqrt{\frac{D_{x-}}{D_{x+}}} \right)^{n}
e^{- 2D_{x} t }I_{n}(2t \sqrt{D_{x+}D_{x-}})
\label{cvd}
\end{equation}
where $D_{x\pm}= D_{x} \pm \frac{1}{2}J_{x}$, and $I_{n}$ is the Bessel function.

Let $y_{n}(t)$ be another variable obeying Eq.~(\ref{xkd}) with
corresponding coefficients $D_{y}$ and $J_{y}$.  Then, the ``overlap''
of the variables, $\sum\limits_{n}x_{n}(t)y_{n}(t)$, can be calculated
as
\begin{equation}
\sum\limits_{n}x_{n}(t)y_{n}(t)= x_{0}y_{0}
e^{- 2(D_{x} + D_{y})t }
I_{0}\left(t'\right) \,,
\label{dvd}
\end{equation}
where $t' =2t \sqrt{(D_{x+} + D_{y-})(D_{x-} + D_{y+})}$.
This expression is used to derive Eqs.~(\ref{wnd}), and (\ref{bnd}).

\subsection{Charge transfer probability distribution}\label{solv2}

The solution to the master equation, Eq.~(\ref{fqd}), for the
probability of $n$-charge transfers, $p_{n}$, which has the structure
of Eq.~(\ref{xkd}), can be read off Eq.~(\ref{cvd}) giving
\begin{equation}
p_{n}(\tau) =
e^{- \tau  + v n}
I_{n}\left( \frac{\tau }{\cosh v}\right)
\label{fvd}
\end{equation}
in the dimensionless variables
\begin{equation}
\tau = 2 Dt
\quad,\quad
e^{v}= \sqrt{\frac{D+ \frac{1}{2}I}{D-\frac{1}{2}I}} \;.
\label{evd}
\end{equation}
With $D$ and $I$ evaluated from Eqs.~(\ref{dfd}), and (\ref{iqd}), the
parameters $D$ and $I$ have the relationship $I/2D = \tanh (V/2T)$, as
required by the fluctuation-dissipation theorem, and the parameter $v$
has the meaning of the dimensionless bias, $v= V/2T$.

The charge expectation values, $\langle n^{k}\rangle= \sum\limits_{n}
n^{k}p_{n}$ and the moments $\langle (\Delta n)^{k} \rangle= $, where
$\Delta n = n - \langle n \rangle$, can be obtained from the
generating functions, $F_{u}$ and ${\cal F}_{u}$,
\begin{equation}
F_{u}(\tau )= 
\sum\limits_{n= - \infty }^{\infty } p_{n}(\tau)e^{un}
\quad,\quad 
{\cal F}_{u} = 
\sum\limits_{n= - \infty }^{\infty } p_{n}(\tau)e^{u(n - \langle n \rangle)}
\label{7pd}
\end{equation}
as
\begin{equation}
\langle n^{k}\rangle = \left.\frac{\partial^{k} F_{u}}
{\partial u^{k}}\right|_{u=0} \;,\;
\langle (\Delta n)^{k}\rangle = \left.\frac{\partial^{k} {\cal F}_{u}}
{\partial u^{k}}\right|_{u=0} \;.
\label{gvd}
\end{equation}
After summation  Eq.~(\ref{7pd}) with $p_{n}$ from
 Eq.~(\ref{fvd}),
we get
\begin{equation}
 F_{u}(\tau ) = \exp\left[\tau \;
\frac{\cosh(v+ u)}{\cosh v}
- \tau  \right] \;,\;
{\cal F}_{u} = 
 e^{- u \tau \tanh v}F_{u}(\tau )
\label{8pd}
\end{equation}

For the first five moments one gets:
$
\langle n \rangle = \tau \tanh v\;,\;
\langle (\Delta n)^2\rangle = \tau \;,\; 
\langle (\Delta n)^3\rangle = \tau \tanh v \;,\; 
\langle (\Delta n)^4\rangle = \tau + 3 \tau^2\;,\; 
\langle (\Delta n)^5\rangle = \tau (1   + 10 \tau )\tanh v
$.
In particular, the third moment satisfies the relation
\begin{equation}
\langle (\Delta n)^3\rangle
 = \langle n\rangle 
\label{2qd}
\end{equation}
in agreement with Ref.~\onlinecite{LevRez01}.


\begin{thebibliography}{99}
\bibitem{vanishG1} 
If $G_{1} \rightarrow 0$, the spin up and down probability is equal,
$u_{n}(t)=  d_{n}(t)$, and the vanishing of the function in
Eq. (\ref{wnd}) 
is no longer exponential in time but decaying as
$1/\sqrt{t}$, a decay purely due to the equal diffusion broadening
of the probability distributions.
\bibitem{ratio} 
In the simplest model where $\hat{T}_{\bm{lr}}= v + w\tau_{z}$ is a
constant, 
the ``constant'' $c$ in Eq.~(\ref{mnd}) equals $c= 
\left(
1 + \frac{2|T_{u}T_{d}| }{|T_{u}|^2+|T_{d}|^2}
 \right) \left(
1 + \sqrt{1 - \left(
\frac{|T_{u}|^2 - |T_{d}|^2}{T_{u}|^2 + |T_{d}|^2}
\tanh \frac{V}{2T}
 \right)^2}
 \right)^{-1}
$, where $T_{u,d}= v\pm w$. For any value of the parameters,
$\frac{1}{2}<c<1$.
\end{thebibliography}
\end{document}